\pdfoutput=1
\documentclass[twocolumn]{aa}  

\usepackage{graphicx}
\usepackage{natbib}
\usepackage[varg]{txfonts}

\newcommand{\fastwind}{{\sc fastwind}}
\newcommand{\cmfgen}{{\sc cmfgen}}
\newcommand{\wmbasic}{{\sc wm-basic}}
\newcommand{\isawind}{{\sc isa-wind}}
\newcommand{\dzero}{\ensuremath{\mathit{D}_{\circ}}}
\newcommand{\ha}{H$\alpha$}
\newcommand{\lsun}{\ensuremath{L_{\odot}}}
\newcommand{\lstar}{\ensuremath{L_{\star}}}
\newcommand{\rstar}{\mbox{$R_{\star}$}}
\newcommand{\mdot}{\ensuremath{\dot{M}}}                      
\newcommand{\msun}{\ensuremath{M_{\odot}}}
\newcommand{\msunyr}{\ensuremath{M_{\odot} {\rm yr}^{-1}}}
\newcommand{\rsun}{\ensuremath{R_{\odot}}}
\newcommand{\teff}{\ensuremath{T_{\rm eff}}}  

\newcommand{\Dmom}{\mbox{$D_{\rm mom}$}}
\newcommand{\vesc}{\mbox{$v_{\rm esc}$}}
\newcommand{\vinf}{\ensuremath{v_{\infty}}}

\newcommand{\kmsec}{\mbox{km\,s$^{-1}$}}
\newcommand{\zsun}{\ensuremath{Z_{\odot}}}

\newcommand{\hii}{\ion{H}{ii}}

\newcommand{\heii}{\ion{He}{ii}}

\newcommand{\civ}{\ion{C}{iv}}
\newcommand{\niv}{\ion{N}{iv}}
\newcommand{\ov}{\ion{O}{v}}
\newcommand{\ovi}{\ion{O}{vi}}
\newcommand{\pv}{\ion{P}{v}}

\begin{document}

\title{The empirical metallicity dependence of the mass-loss rate \\ of O- and
       early B-type stars}   

\authorrunning{M.\,R.~Mokiem et al.}
\titlerunning{The emperical \mdot(Z) dependence of O and early-B stars}

\author{M.\,R.~Mokiem\inst{1}
   \and A.~de~Koter\inst{1}
   \and J.\,S.~Vink\inst{2}
   \and J.~Puls\inst{3}
   \and C.\,J.~Evans\inst{4}
   \and S.\,J.~Smartt\inst{5}
   \and P.~A.~Crowther\inst{6}
   \and A.~Herrero\inst{7,8}
   \and N.~Langer\inst{9}
   \and D.\,J.~Lennon\inst{10,7}
   \and F.~Najarro\inst{11}
   \and M.\,R.~Villamariz\inst{12,7}
}


\institute{ 
  Astronomical Institute Anton Pannekoek, University of Amsterdam,
  Kruislaan 403, 1098~SJ Amsterdam, The Netherlands
  \and 
  Astrophysics Group, Lennard-Jones Laboratories, Keele University,
  Staffordshire, ST55BG, UK
  \and 
  Universit\"ats-Sternwarte M\"unchen, Scheinerstr. 1,
  D-81679 M\"unchen, Germany
  \and 
  UK Astronomy Technology Centre, Royal Observatory, Blackford Hill,
  Edinburgh, EH9 3HJ, UK
  \and 
  Astrophysics Research Centre, School of Mathematics and Physics,
  Queen's University Belfast, Belfast BT7 1NN, UK
  \and
  Department of Physics and
  Astronomy, University of Sheffield, Hicks Building, Hounsfield Rd,
  Shefffield, S3 7RH,  UK
  \and
  Instituto de Astrof\'{\i}sica de Canarias, E-38200, La Laguna,
  Tenerife, Spain
  \and
  Departamento de Astrof\'{\i}sica, Universidad de La Laguna,
  Avda.\ Astrof\'{\i}sico Francisco S\'anchez, s/n, E-38071
  La Laguna, Spain
  \and
  Astronomical Institute, Utrecht University, Princetonplein 5,
  3584 CC Utrecht, The Netherlands
  \and
  The Isaac Newton Group of Telescopes,
  Apartado de Correos 321, E-38700,
  Santa Cruz de La Palma, Canary Islands, Spain
  \and 
  Instituto de Estructura de la Materia, Consejo Superior de
  Investigaciones Cient\'{\i}ficas, CSIC, Serrano 121, E-28006
  Madrid, Spain
  \and
  Grantecan S.A., E-38200, La Laguna,
  Tenerife, Spain
}

\date{Accepted version}

\abstract{We present a comprehensive study of the observational dependence
of the mass-loss rate in stationary stellar winds of hot massive stars on
the metal content of their atmospheres.  The metal content of stars in the
Magellanic Clouds is discussed, and a critical assessment 
is given of
state-of-the-art mass-loss determinations of OB stars in these two satellite
systems and the Milky-Way.  Assuming a power-law dependence of mass loss on
metal content, $\mdot \propto Z^{m}$, and adopting a theoretical relation
between the terminal flow velocity and metal content, $\vinf \propto
Z^{0.13}$ (Leitherer et al.\ 1992), we find $m = 0.83 \pm 0.16$ for 
non-clumped outflows from an analysis of the wind momentum luminosity relation 
(WLR) for stars more luminous than $10^{5.2} \lsun$. Within the 
errors, this result is in agreement with the prediction $m = 0.69 \pm 0.10$ by
Vink et al.\ (2001). Absolute empirical values for the mass loss, based on
\ha\ and ultraviolet (UV) wind lines, are found to be a factor of two higher
than predictions in this high luminosity regime. If this difference is
attributed to inhomogeneities in the wind, and this clumping 
does not impact the predictions,
this would imply that luminous O and early-B stars have clumping factors 
in their \ha\ and UV line forming regions of about a factor of four.
For lower luminosity stars, the winds are so weak that their strengths can 
generally no longer be derived from optical spectral lines (essentially \ha) 
and one must currently rely on the analysis of UV lines.
We confirm that in this low-luminosity domain the observed Galactic WLR is
found to be much steeper than expected from theory (although the specific
sample is rather small), leading to a discrepancy
between UV mass-loss rates and the predictions by a factor 100 at
luminosities of $L \sim 10^{4.75} \lsun$, the origin of which is
unknown. We emphasize
that even if the current mass-loss rates of hot luminous
stars would turn out to be overestimated as a result of wind clumping, 
but the degree of clumping would be rather independent of metallicity, the
scalings derived in this study are expected to remain correct. }

\maketitle


\section{Introduction}

Understanding the properties and evolution of massive stars in low
metallicity environments is of fundamental importance in astrophysics. This
is amply illustrated by referring to the anticipated role of massive stars
in the early universe, where they are thought to be responsible for its
re-ionization \cite[e.g.][]{haehnelt01, wyithe03}, galaxy formation, and
chemical evolution of young galaxies and of the intra-galaxy medium. Strong
indications of gamma-ray bursts occurring primarily at low metallicity
\cite[e.g.][]{gorosabel05} and massive stars being progenitors
\cite[e.g.][]{hjorth03}, adds to the point. To a large extent, the
evolution of massive stars is controlled by the effects of mass loss,
therefore mass loss has a direct and/or indirect impact on all the phenomena
mentioned above. This makes a quantitative handle on the dependence of mass
loss on the metal content of the environment out of which these early stars
form so vital. Obviously this argument also applies to massive stars in the
local universe.

In the last decade the development of realistic stellar atmosphere models
has allowed accurate determination of the wind parameters of a
sizeable sample of early-type Galactic stars \citep[e.g.][]{puls96,
herrero02, repolust04, crowther06}. This, in turn, has allowed us to test
our understanding of aspects of stellar wind hydrodynamics, including its
driving mechanism, to an unprecedented level. 
During this period new instrumentation developements on large telescopes
have resulted in significant numbers of extragalactic stars being
spectroscopically studied. Examples of extensive surveys are those of
Massey et al. (2004, 2005) and of Evans et al. (2005), although there
are many others and the relevant ones will be discussed later.
The latter
program is the {\em VLT-FLAMES Survey of Massive Stars}, in which close to
100 O- and early B-type stars were observed in the Magellanic Clouds.
The inclusion of line blocking/blanketing in the modeling of hot star
atmospheres \citep[e.g.][]{hubeny95, hillier98} allows for a consistent
description of the effects from vast numbers of overlapping metal lines,
particularly regarding a detailed treatment of the effects of chemical
composition, and has paved the way for a robust quantitative comparison of
the wind strengths of hot massive stars in different metallicity
environments. The most recent step forward in the modeling technique has
been the development of an automated fitting method, opening up a means to
analyse large samples in a homogeneous way \citep{mokiem05}.

This paper provides a critical assessment 
of the current standing of mass-loss
determinations in the Galaxy and the Large and Small Magellanic Clouds,
and is specifically aimed at establishing the dependence of mass loss
on the chemical abundance pattern, notably metal content $Z$.  The Small
Magellanic Cloud (SMC) with $Z$ at about 1/5th of the Galactic value
is the lowest metallicity we can so far probe in reasonable detail.
The Large Magellanic Cloud (LMC) with approximately half the metal
content of the Galaxy provides an intermediate environment.

The only possibility, in the forseeable future, of studying
stars in environments with significantly lower metallicity
than the SMC is the massive stellar populations of the Local Group
dwarfs GR8, Sextans A and Leo A. These three galaxies have
metallicities betweem 0.1 and 0.05\,Z$_{\odot}$, but their low
starformation rates and distances of around 1-2\,Mpc makes studies
of large numbers of young massive stars extremely difficult. Given this 
difficulty of establishing the 
dependence of mass-loss rates upon metallicity below 1/5
Z$_{\odot}$, 
we must rely on predictions to access this interesting part of parameter 
space. Although we do not deal with the theory of wind driving mechanism, 
we will compare predicted and observed wind strengths to establish 
successes and failures of the theory such that we may identify aspects 
of the $\mdot(Z)$ problem that require further study, and regimes in
parameter space that seem sufficiently under control to allow us to
venture out to lower metallicity regimes and the early universe 
\citep[see also][]{vink05}.

In Sect.~\ref{sec:mdotmechanism} we briefly review the mass-loss
mechanism of early-type stars. The present day chemical composition of
the Galaxy and Magellanic Clouds is discussed in
Sect.~\ref{sec:abundances}. Sections~\ref{sec:mdot_diag} and
\ref{sec:mdot_comp} describe the methodology used to determine 
mass-loss rates and how to compare these for different chemical 
environments. The
observed mass-loss relations are presented in Sect.~\ref{sec:obs_wlr}.
On these relations the global metallicity dependence of \mdot\
is determined in Sect.~\ref{sec:mdot_vs_z}. Finally, in
Sect.~\ref{sec:disc} we discuss the observed mass-loss relations in
terms of the so-called weak wind problem and wind clumping, and end
with concluding remarks.

\section{The mass-loss mechanism of early-type stars}
\label{sec:mdotmechanism}

We briefly review those aspects of the wind driving mechanism of
massive early-type stars that are relevant in the context of
establishing an empirical relation between mass loss and chemical
composition. For a more in-depth treatment of the physics of mass
loss, see e.g. \cite{kudritzki00} and \cite{vink01}. The basic
mechanism driving the winds of hot massive stars is the transfer of
momentum from photons to the atmospheric gas by line interactions. The
driving mechanism implies that the properties of the stellar wind will
depend on the number of photons per second streaming through the
photospheric layers (reflecting the stellar luminosity), and on the
number and ability of lines being available -- in particular at
wavelengths around the photospheric flux maximum -- to absorb or 
scatter these photons.

\begin{table}[t]
\caption{Overview of the ions that dominate the line driving in
  early-type stars -- for solar metallicity with
  $\lstar = 10^{5}$~\lsun. See Sect.~\ref{sec:mdotmechanism} for a discussion. 
}
\begin{small}
\begin{center}
\begin{tabular}{lrrrr}
\hline\\[-9pt] \hline\\[-7pt]
         & {\teff = 20\,kK} & {\teff = 25\,kK} & {\teff = 30\,kK} & {\teff = 40\,kK} \\
 Atom & Contr. \%        & Contr. \%        & Contr. \%        & Contr. \% \\
\hline\\[-9pt]
 H    & 19.53  & 14.84 & 12.82 & 14.11  \\
 He   &  0.01  &  0.03 &  0.05 &  0.05  \\
 C    &  2.61  & 12.28 & 14.43 & 24.03  \\
 N    &  0.49  &  1.09 &  1.76 &  7.19  \\
 O    &  0.65  &  0.14 &  1.87 &  4.64  \\
 Mg   &  2.07  &  0.46 &  3.08 &    --  \\
 Al   &  2.17  &  3.21 & 12.61 &    --  \\
 Si   &  7.95  & 10.01 & 20.34 & 13.16  \\
 P    &  6.56  &  9.46 &  0.93 &  1.13  \\
 S    &  4.51  &  6.52 &  4.55 & 16.51  \\
 Cl   &  0.30  &  1.57 &  1.05 &  0.91  \\
 Ar   &  0.11  &  0.24 &  0.33 &  1.87  \\
 Cr   &  2.36  &  1.98 &  0.78 &  0.48  \\
 Mn   &  1.10  &  0.92 &  1.07 &  0.34  \\
 Fe   & 39.00  & 32.29 & 23.07 & 10.51  \\
 Ni   &  6.40  &  4.03 &  0.65 &  2.40  \\
 Rest &  4.18  &  0.93 &  0.61 &  2.67  \\
\hline
\end{tabular}
\end{center}
\end{small}
\label{tab:line_force_contribution}
\end{table}

The dependence of the wind driving on the number of lines present suggests
that mass loss is a function of elemental abundance. Whether this is indeed
the case formally depends on the nature of the driving lines. In the
hypothetical case that one would be in a regime of abundances for which all
lines effectively contributing to the line force are optically thick, mass
loss would {\em not} be a function of elemental abundance. This regime is
not encountered in even the most metal rich environments known
\citep{vink01}. In reality, it is found that the lines driving the wind are
a mixture of optically thin and optically thick lines. Representing the
distribution of line strengths by a power law, one predicts, for
Galactic O stars, a ratio of the line acceleration from optically thick lines
to the total line acceleration of $\alpha \sim 2/3$ \citep[see][]{puls00},
where an even larger contribution by optically thick lines would increase
this value (and vice versa). This ensures a dependence of mass loss on
elemental abundance.

Which elements dominate the line force? The answer depends on the effective
temperature of the star. \citep[More precisely: on the radiation and
electron temperature in the wind acceleration regime; see
also][]{vink99,puls00}.  Although hydrogen and helium are by far the most
abundant elements, their impact on the wind driving is modest. Decisive for
whether or not a species is a significant contributor to the line driving is
the product: abundance $\times$ ionisation fraction $\times$ number of
effective lines. Very roughly, the elemental abundance times the ionisation
fraction of hydrogen and helium are similar to that of metals that are in
the dominant stage of ionisation. It is therefore the {\it number} of
driving lines that is decisive. As H and He -- due to their simple atomic
structure -- have only few lines that can effectively contribute to the line
force, it is relatively abundant complex atoms that are the main
contributors.

We estimate the relative contributions of the different elements
by the use of a Monte Carlo method \citep []
{abbott85,dekoter97,vink99} that calculates the total momentum
transfer from the radiation field to the outflowing gas
particles. Although the relative contribution to the line force is
depth-dependent (see below), here we simply register the atomic number of the
elements with which the photons interact somewhere in the wind, and we
present the results in Tab.~\ref{tab:line_force_contribution}.  For a
late-O dwarf of solar composition, CNO accounts for some 15 percent;
iron contributes some 25 percent.  Other iron-group elements (for
instance Cr, Mn, Co, Ni) add a few percent.  $\alpha$-elements, such
as Ne, Mg, Si, S, Ar, and Ca account for about 30 \%, with Si being
the main contributor with $\sim$ 20 \%.  We mention a decomposition of
the line force in terms of iron group and $\alpha$ elements as
their nucleosynthetic origin is different. 
The former are mostly produced in thermonuclear Type Ia supernovae,
the latter predominantly in core-collapse supernovae of types II and Ib/c.
Beware that due to their different electronic structure specific groups
of elements have a different line-strength statistics, and dominate
the line acceleration at different depths: iron group elements have
a somewhat larger influence on the mass-loss rate, whereas lighter ions
dominate the acceleration in the outer wind, thus controlling the terminal
velocity (cf. \citealt{vink99, puls00}). These more subtle effects are
not reflected in the statistics provided in 
Table~\ref{tab:line_force_contribution}.

Mixing of CNO-cycled material to the surface during the supergiant
phase may affect the relative abundances of these three
elements. However, in terms of their contribution to the line force
not much will change when this happens, as these three elements have
more or less equal numbers of effective driving lines near the
photospheric flux maximum and the C+N+O abundance remains unaffected
by the CNO-cycle \citep{vink02}.

Accounting for the fact that we are interested in the gross dependence
of wind parameters on metallicity, and particularly in the product of
mass-loss rate and terminal velocity (see below),
we conclude that {\em a straight mean of C+N+O, $\alpha$-elements, and
iron-group elements is an appropriate abundance value to use as a $Z$
measure for O stars.}  For stars of spectral type B and A 
the contribution of iron increases to up to 50 percent.

\section{Metallicity determinations of early-type stars}
\label{sec:abundances}

Having identified the elements that dominate the line force, we now
address the present-day chemical composition of hot massive stars in
the Galaxy and the Magellanic Clouds.

Spectroscopy of \hii\ regions and supernova remnants have been used to
probe the gas phase composition of the Magellanic Clouds \citep[for
reviews see for instance][]{pagel78, dufour90, russell90, garnett99}.
The results derived from these studies do not necessarily represent
the initial chemical composition of the stars in these environments
because of fractionation of some elements from the gas phase onto
solid state particles (or ``dust''). Also, emission line studies may
be affected by spatial inhomogeneities in the nebula.
Therefore, to obtain the appropriate chemical composition of the winds we 
are studying it would be pertinent to determine the photospheric abundances
of the massive stellar populations in the Clouds.
As the comparison between theoretical and observed
wind strength is usually not done on an individual basis \citep[but
see][]{dekoter97} it is relevant to specify how the observed
metal abundance $Z$ -- so far the abundance input parameter for 
mass-loss predictions -- is derived. If $Z$ is derived from CNO abundances
in either evolved or rapidly rotating objects one should be careful.
In supergiants and even giants \citep{korn00,
lennon03} rotational mixing and ejection of the outermost
envelope by mass loss may bring CNO equilibrium material to the
surface. In stars rotating at about half of break-up at the
surface or more, abundance alterations due to rotational mixing may already
occur very early on in the stars evolution
\citep{yoon05}.

The optical spectra of O stars show few spectral lines of heavy
elements. B stars are relatively rich in absorption features due
to C, N, O, Mg, Al, Si, S, and Fe providing a sound basis for establishing
the metallicity, although we note that the Fe lines are 
intrinsically quite weak \citep[especially so in the
SMC;][]{rolleston03}. The advantage of B and later-type dwarf and
giant stars
is also that they may be analysed accurately using non-LTE line-blanketed
hydrostatic atmospheres, without winds.

\begin{table*}[t]
\caption{Present-day chemical composition of the LMC and SMC from 
B-type stars. The values are taken from \cite{hunter06} and Trundle
et al. (2007) apart from Al and S  which are LTE results from 
\citet{rolleston02,rolleston03}.  For comparison
the solar abundances of \cite{asplund05} also given.}
\begin{small}
\begin{center}
\begin{tabular}{lrrrrrr}
\hline\\[-9pt] \hline\\[-7pt]
  Element & Solar       & \multicolumn{2}{c}{LMC}           & \multicolumn{2}{c}{SMC} \\
        &             & $12+\log X/{\rm H}$ & $\Delta [X/{\rm H}]$  & $12+\log X/{\rm H}$ & $\Delta [X/{\rm H}]$ \\
\hline\\[-9pt]
 C    &  8.39  &  7.73 &  $-$0.66 &   7.37 &  $-$1.02  \\
 N    &  7.78  &  6.88 &  $-$0.90 &   6.50 &  $-$1.28  \\
 O    &  8.66  &  8.35 &  $-$0.31 &   7.98 &  $-$0.68  \\
 Mg   &  7.53  &  7.06 &  $-$0.47 &   6.72 &  $-$0.81  \\
 Al   &  6.37  &  ...  &  ...     &   5.43 &  $-$0.72  \\
 Si   &  7.51  &  7.19 &  $-$0.32 &   6.79 &  $-$0.72  \\
 S    &  7.14  &  ...  &  ...     &   6.44 &  $-$0.42  \\
 Fe   &  7.45  &  7.23 &  $-$0.29 &   6.93 &  $-$0.57  \\
\hline
\end{tabular}
\end{center}
\end{small}
\label{tab:abundances}
\end{table*}

A major goal of the {\em VLT-FLAMES survey of massive stars} \citep{evans05} 
was to observe a large number of such B-type stars in the same vicinity as the 
O-stars to determine their photospheric abundances. 
\cite{hunter06} and \cite{trundle07} have presented the chemical compositions
of 34 and 53 B-type stars in the SMC and LMC respectively. These are the 
narrow lined stars from the FLAMES survey, with the highest signal-to-noise
spectra, analysed with the line-blanketed non-LTE code {\sc tlusty}.
These authors determined the best estimates for the chemical 
composition of the Clouds from a comparison of results from stellar and 
nebular work, and we summarise and extend their results in 
Table\,\ref{tab:abundances}. The non-LTE absolute abundances for 
C, N, O, Mg, and Si are quoted, along with the simple difference between
these absolute values and the current standard solar abundances of 
\cite{asplund05}. 
However detailed non-LTE model atoms for use in calculating line 
profiles for Al, S and Fe in {\sc tlusty} are required to allow reliable
line abundances to be calculated consistently in non-LTE.
The results for Fe
are taken from \cite{trundle07} in which a non-LTE atmosphere
and LTE line formation calculations were used. The results for Al and S 
are taken from the LTE B-star analysis of 
\cite{rolleston03} and \cite{rolleston02}. Because we are less certain
of the validity of the absolute values of the LTE results, we quote the
differential abundances from these papers which were determined from 
analysing Galactic B-stars with identical atmospheric parameters. 
We should also note that the \cite{rolleston03} abundances come from the 
analysis of a single star AV304, and the LMC results are from 5 stars 
with spectra of modest signal-to-noise.

The SMC abundance of the $\alpha$ elements (O, Mg, Si) has a mean relative to 
solar of $\Delta [\alpha / {\rm H}] = -0.7 \pm 0.1$~dex. The differential
LTE result of Al is in good agreement with this value, although the one S abundance
from the star AV304 is somewhat higher. The Fe abundance is 
$\Delta [{\rm Fe}/{\rm H}] = -0.57 \pm 0.16$~dex, which, 
within the uncertainties, is in agreement with the $\alpha$ elements. 
The SMC Fe abundance has also been determined through studies of AFGK
supergiants, which have more and stronger metal lines in their
spectra and for which non-LTE effects are less important. 
They tend to show very good agreement at $\Delta [{\rm Fe}/{\rm H}] = -0.60 \pm 0.1$~dex
\citep{venn1999}. Differences in modelling techniques/codes,
diagnostic lines, and atomic data will clearly lead to systematic differences
between the sets of results. The discrepancies between different analyses are becoming
smaller, but have not completely been resolved and progress is required in 
the model atmosphere and line formation codes, particularly with regard to atomic data
for S, Al and Fe for B-star analysis. We adopt $\Delta [Z/{\rm H}] \sim
-0.7\pm0.1$ for the SMC, but will discuss the effects of a potentially different
metallicity later (see Sect.~\ref{sec:empirical-dmdtz}).

The abundance of the $\alpha$ elements in the LMC relative to solar 
from \cite{hunter06} and \cite{trundle07} is $\Delta [\alpha / {\rm H}] = -0.36 \pm 0.1$~dex. 
An extensive comparison of abundance determinations in the LMC was compiled by
\cite{rolleston02}. Reasonable agreement is found between six
independent studies of the overall metallicity of the LMC --
irrespective of the class of object used to trace the chemical
composition or the spatial location of the investigated stars. The
mean metallicity, based on O and Si, is ($\Delta [{\rm O}/{\rm H}] +
\Delta [{\rm Si}/{\rm H})/2 = -0.30 \pm 0.08$~dex. Some previous 
analyses using OB stars seem to indicate that iron is less
deficient than the $\alpha$-elements. \cite{rolleston02} report an $\alpha$ to iron ratio
$\Delta [\alpha / {\rm Fe}] \sim -0.14$~dex for main sequence OB
stars, consistent with that obtained for evolved B-stars
\citep{korn00}. However \cite{trundle07} determine a mean 
differential Fe abundance of $-0.29\pm0.15$ for 13 stars of their
sample in NGC2004, in good agreement with the depletions of O, Mg and Si. 
Studies using F supergiants and
Cepheids, however, report iron abundances representative of the mean
metallicity \citep{hill95, luck98}.
In this study, we adopt $\Delta [Z/{\rm H}] \sim -0.3 \pm 0.1$ 
for the LMC.

We note that for C and N the values in \cite{hunter06} are the 
best estimates of the {\em baseline} abundances in the Clouds. 
I.e. they are estimates of the C and N when the stars are 
born and before any rotational mixing has caused the photospheric
abundances to change. It has been clear for some time that C and 
N are significantly underabundant in the 
Clouds in comparison to the heavier elements. This may have an 
effect on the wind strengths of the hottest stars where 
CNO (mainly C) is a significant contributor to the line force. 
(see Table~\ref{tab:line_force_contribution}).

\section{Mass-loss determinations of early-type stars}
\label{sec:mdot_diag}

During recent years, three basic diagnostics to derive the mass-loss
rate of early-type stars have been employed. These are: 
(a) infrared, millimetre and radio excess due to free-free
    processes;
(b) ultraviolet resonance lines, and 
(c) optical lines, notably \ha, but for the hottest stars also 
    \heii\,$\lambda 4686$.
For a didactic explanation of the way in which \mdot\
can be derived from these diagnostics, see
\cite{lamers99}. In the context of the dependence of mass
loss on metal content, millimetre and radio excess emission in
early-type stars is (at this moment) impracticable as the flux levels
are weak and can only be measured for not too distant Galactic stars
(up to a few kiloparsec), i.e. the method can not be applied to
Magellanic Cloud stars.

UV resonance lines of relatively abundant elements such as C, N, O and
Si are the most sensitive probes of mass loss, allowing detection of
rates as low as $\sim 10^{-9}\,\msunyr$. The lines typically saturate
at about $\sim 10^{-7}$\,\msunyr, therefore for strong winds they only
provide lower limits to \mdot. Because of the ease with which the
lines saturate, unsaturated lines often relate to minor ionisation
stages (and/or, obviously, to weak winds). The ionisation of these
trace ions may depend rather critically on the treatment of
line-blanketing, clumping, and shocks. Over the past decade, line-blanketing 
has been incorporated into model atmosphere codes for hot stars with 
winds. In a detailed comparison of three of these codes, \cite{puls05}
conclude that the flux levels at $\lambda \gtrsim 400$ \AA\ agree very
well. Below 400 \AA, discrepancies are found, implying that the
population of species such as \niv, \civ, and \ovi\ should be taken 
with care, as should the mass-loss rates derived using these ions.
One of the most important current challenges in model atmospheres
treating outflows is to implement the direct and indirect effects of
the line-driven instability \citep[see e.g.][and references
therein]{lucy70,owocki94}. This instability causes the formation of
small scale density and velocity gradients in the wind flow, creating
``clumps'' of gas and shocks leading to X-ray emission and enhanced
EUV-flux. We will discuss clumping in more detail in
Sect.~\ref{sec:clumping}. Shocks may also impact on the ionisation of
the species mentioned above. This presents a second reason for being
cautious with the \mdot\ values derived from UV resonance lines of
which the ionisation is affected by EUV and/or X-ray photons. A third
reason to be cautious is that the abundances of C, N, and O may be
affected by the surfacing of nuclear processed material (see
Sect.~\ref{sec:abundances}).

The third approach to derive the rate of mass loss is by fitting the \ha\
profile, and, for stars of roughly spectral type O5 or earlier,
\heii\,$\lambda 4686$. The advantages of fitting these lines compared to
ultraviolet resonance lines are that {\em i)} the abundance determination is
robust \citep[for the most detailed approach, see][]{mokiem05, mokiem06};
{\em ii)} at least for hydrogen, the ionisation balance is well known,
i.e. it is not affected by shock processes (which might still be a
problem for helium). On the down side however, the \ha\ diagnostic is not
as sensitive as are the UV lines. For OB stars, only for mass-loss rates in
excess of $\sim 10^{-7}$ \msunyr\ can the contribution of wind
emission to the \ha\ photospheric line
be used to determine \mdot. Even more, at such low wind
densities the uncertainty about the velocity law (which cannot be
constrained if the wind emission inside \ha\ is low) introduces an additional
ambiguity, which increases the errors in the derived
mass-loss rate considerably, up to factors of two to three (e.g.,
\citealt{puls96}).

This poses a challenge: {\em the \ha\ and {\rm UV} resonance line
diagnostics have only a small mass-loss regime in common in which both
techniques can be applied simultaneously}.  This is unfortunate in view of
resolving the weak wind problem (see Sect.~\ref{sec:weakwinds}), although
there may be alternative wind diagnostics. 
First, the Br$\alpha$ line at 4.05\,$\mu$m is
intrinsically stronger than \ha\ and could push the sensitivity of the
hydrogen line \mdot-diagnostic by possibly a factor two to three
\citep{schaerer96, lenorzer04, repolust05} or even more (as
resulting from test calculations by JP and FN), providing sufficient
overlap to derive \mdot\ from both methods for a reasonable number of stars.
Second there is a \pv\ resonance doublet at
$\lambda\lambda$1118,1128~\AA. The
abundance of phosphorus is about a factor of $10^{2}-10^{3}$ less than that
of C, N, O, and Si. As a result of this the line does not saturate as
easily. Consequently, it can be applied to stars with \mdot\ in excess of
$10^{-7}$~\msunyr\ \cite[see e.g.][]{crowther02,hillier03}. Moreover, for 
a range of O subtypes \pv\ is expected to
be the dominant ionisation stage (for recent results, see
\citealt{puls07}), making it less susceptible to shocks.
Access to the far-ultraviolet region of the spectrum is provided by, for
instance, the {\em Far Ultraviolet Spectroscopic Explorer}. Analysis of
spectra obtained with this instrument strongly points to discrepant P\,{\sc
v} and \ha\ based mass-loss rates, suggestive of clumped winds
\citep{massa03, fullerton06}.

\section{Comparing \mdot\ in different environments}
\label{sec:mdot_comp}

How should we compare the mass-loss rates of OB stars in different
galaxies?  The most straightforward way of doing this would be
to consider two stars with almost identical parameters -- one in each
galaxy -- and directly compare their \mdot\ values.  There are
two arguments against such an approach. The first one is of a
practicable nature. The number of O and early-B stars studied in
the Galaxy and the Magellanic Clouds is so far too limited to identify
a significant number of stars with ``identical'' luminosity,
temperature, mass, and to a lesser extent rotational and terminal wind
velocity. The impact of stellar rotation on mass loss appears
important only for stars close to the $\Omega$-$\Gamma$ limit
(e.g. \citealt{maeder00} and \citealt{lamers04}, and, e.g., 
\citealt{vanboekel03} and \citealt{smith03} for the extreme case of 
$\eta$\,Carinae), therefore,
for relatively ``normal'' early-type stars one may consider it less
critical. Terminal velocity differences due to metallicity effects are
found to be minor, both from observational \citep[e.g.][]{evans04a}
and theoretical \citep[e.g.][]{puls00, krticka06} considerations.

Even if for a few cases one could perform such a direct comparison, it would
still not be an appealing approach, as the $\mdot(Z)$
relation may not necessarily be universal for all stellar parameters, and 
the result might not be applicable to all early-type stars.  
Therefore, the second reason against an object-to-object comparison is that it
may ignore potential
physical arguments that would allow the entire parameter space of hot massive 
stars be used to establish the mass loss-metallicity relation, and that would add 
predictive power to the derived $\mdot(Z)$.

Indeed the radiation-driven wind theory makes such predictions, and 
a powerful way to proceed is through the use of the so-called
{\em modified wind momentum -- luminosity relation},
\citep[WLR; e.g.][]{kudritzki95,kudritzki00}
\begin{equation}
   \log \Dmom \equiv \log \left( \mdot \vinf \sqrt{R}\right) 
          \simeq x \log (\lstar / \lsun) + \log \dzero~.
\end{equation}
In this relation the slope $x$ and the constant \dzero\ are
expected to vary as a function of spectral type and metal content
\citep{kudritzki99,vink00,puls00}.
This equation expresses that the mechanical momentum of the stellar
wind flow is primarily a function of photon momentum. It is perhaps
surprising that the stellar mass does not feature in this
dependence. The reason is that 
$\log \dzero$ contains a term 
$\propto (3/2-x) \log M$, which (almost) vanishes since $x$
happens to be $\sim 3/2$ for O-stars and early B-supergiants
\citep{puls00}. 
The uniqueness of the modified wind momentum relation 
has been confirmed by independent investigations
\citep[e.g][]{vink00, pulsIAUS212}. In Tab.~\ref{tab:wlr-par}
coefficients for the WLR as predicted by \cite{vink00, vink01} are
given for the Galactic, LMC and SMC metallicities.

\begin{table*}
\caption{Coefficients describing empirical modified-wind momentum
  relations.  Slope and vertical offset are given as $x$ and \dzero,
  respectively.   Parameters denoted with an apostrophe are derived from the
  wind momentum distributions including clumping corrected \Dmom\
  values for stars with \ha\ in emission. For comparison, the coefficients of the theoretically predicted 
  relation of \cite{vink00} is also given. 
}
\begin{small}
\begin{center}
\begin{tabular}{llcccc}
  \hline\\[-9pt] \hline \\[-7pt]
  Galaxy & sample & $x$ & $\log D_0$ & $x'$ & $\log D'_0$\\[1pt]
 \hline \\[-9pt]
  MWG & \cite{mokiem05}   & $1.86 \pm 0.20$ & $18.71 \pm 1.16$ & $1.58 \pm 0.19$ & $20.16 \pm 1.11$\\[3.5pt]
      & Total             & $1.84 \pm 0.17$ & $18.87 \pm 0.98$ & $1.56 \pm 0.16$ & $20.23 \pm 0.91$\\[3.5pt]
      & \cite{martins05b} & $3.15 \pm 0.95$ & $10.29 \pm 5.08$ \\[9pt]
      & \cite{vink00}     & $1.826 \pm 0.044$ & $18.68 \pm 0.26$ \\[9pt]
  LMC & \cite{mokiem07}   & $1.87 \pm 0.19$ & $18.30 \pm 1.04$ & $1.49 \pm 0.18$ & $20.40 \pm 1.00$\\[3.5pt]
      & Total             & $1.96 \pm 0.16$ & $17.88 \pm 0.91$ & $1.57 \pm 0.15$ & $20.02 \pm 0.84$\\[9pt]
  SMC & \cite{mokiem06}   & $2.00 \pm 0.27$ & $17.31 \pm 1.52$ & $1.50 \pm 0.23$ & $20.03 \pm 1.32$\\[3.5pt]
      & Total             & $1.84 \pm 0.19$ & $18.20 \pm 1.09$ & $1.62 \pm 0.19$ & $19.26 \pm 1.10$\\[6.5pt]
\hline\\[7pt]
\end{tabular}

\end{center}
\end{small}
\label{tab:wlr-par}
\end{table*}

Assuming the mass loss and terminal velocity are power laws of
metallicity, i.e. 
\begin{equation}
  \mdot \propto Z^{m}
\end{equation}
and
\begin{equation}
\label{eq:vinf_z}
  \vinf \propto Z^{n}~,
\end{equation}
it follows that 
\begin{equation}
\label{eq:z_dep}
(m+n) = \Delta \log \Dmom / \Delta \log Z
\end{equation}
The index $m$ may now be derived from a comparison of the modified
wind momentum for different galaxies, adopting for instance the
theoretical result $n = 0.13$ from \cite{leitherer92} to describe the
$\vinf(Z)$ dependence. As the slope $\alpha$ of the WLR is not
identical (though similar) for different metallicities one can not
simply substitute the constant $\dzero(Z)$ for \Dmom\ in the above
equation. What is required is a comparison of the actual \Dmom\ at
a chosen luminosity, using the uncertainty in the WML relation at
that specific \lstar\ (see Sect.~\ref{sec:empirical-dmdtz}).

\section{Observed mass-loss relations}
\label{sec:obs_wlr}

To determine the empirical mass loss versus metallicity dependence, we
compiled mass-loss rates and terminal velocity determinations
from the literature, limiting ourselves to results obtained with
state-of-the-art modeling techniques using unified non-LTE
line-blanketed stellar atmosphere models. Initially, we collected
results obtained with four such codes: \fastwind\ of \cite{puls05},
\cmfgen\ of \cite{hillier98}, \wmbasic\ of \cite{pauldrach01}, and
\isawind\ of \cite{dekoter93, dekoter97}.  We decided not to use
studies performed with the latter two computer programmes
\citep[e.g.][]{dekoter94, dekoter98, bianchi02, garcia04} as these
account for only a relatively minor fraction of the total number of
stars investigated (some 20 percent) of which some have been
reanalysed with either \fastwind\ or \cmfgen.  This approach assures
an extensive {\em and} relatively homogeneous dataset, but not to the
limit that only one code is used. In this way we can still investigate
potential (systematic) differences between at least two methods.
Objects for which the \fastwind\ and \cmfgen\ studies provided
only upper limits on the mass-loss rate are not included in the
determination of the empirical WLRs, however they are included
in the figures for reference and comparison purposes. We also limit
the sample to stars with $\teff > 24$~kK. For lower effective
temperatures, a decrease in the terminal wind velocity is
observed \citep []{lamers95,crowther06}, which may be accompanied
by an increase in the mass-loss rate due to a change in the ionization
balance of iron \citep{vink99},
potentially leading to a different WLR for such relatively cool stars.
This excludes some targets from the studies by \cite{trundle04,
trundle05, crowther06}. Finally, we did not include the entries
from the comprehensive study of \cite{markova04} because, with the
exception of the mass-loss rates, the stellar parameters were based on
calibrations. We note however that the stellar parameters derived by
these authors compare well with other studies.

\subsection{Galaxy}\label{sec:gal}

\begin{figure*} 
\begin{center}
\resizebox{18cm}{!}{
\includegraphics{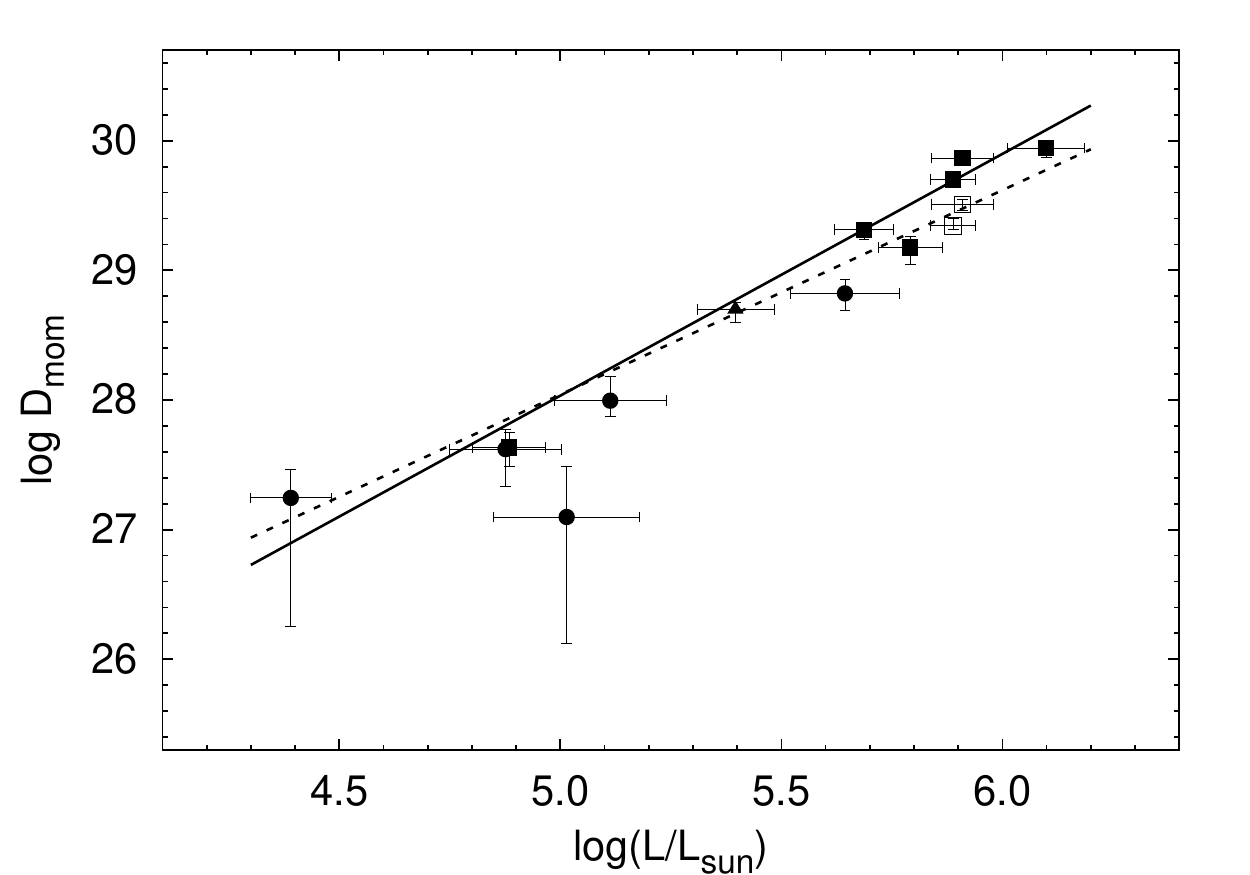}
\includegraphics{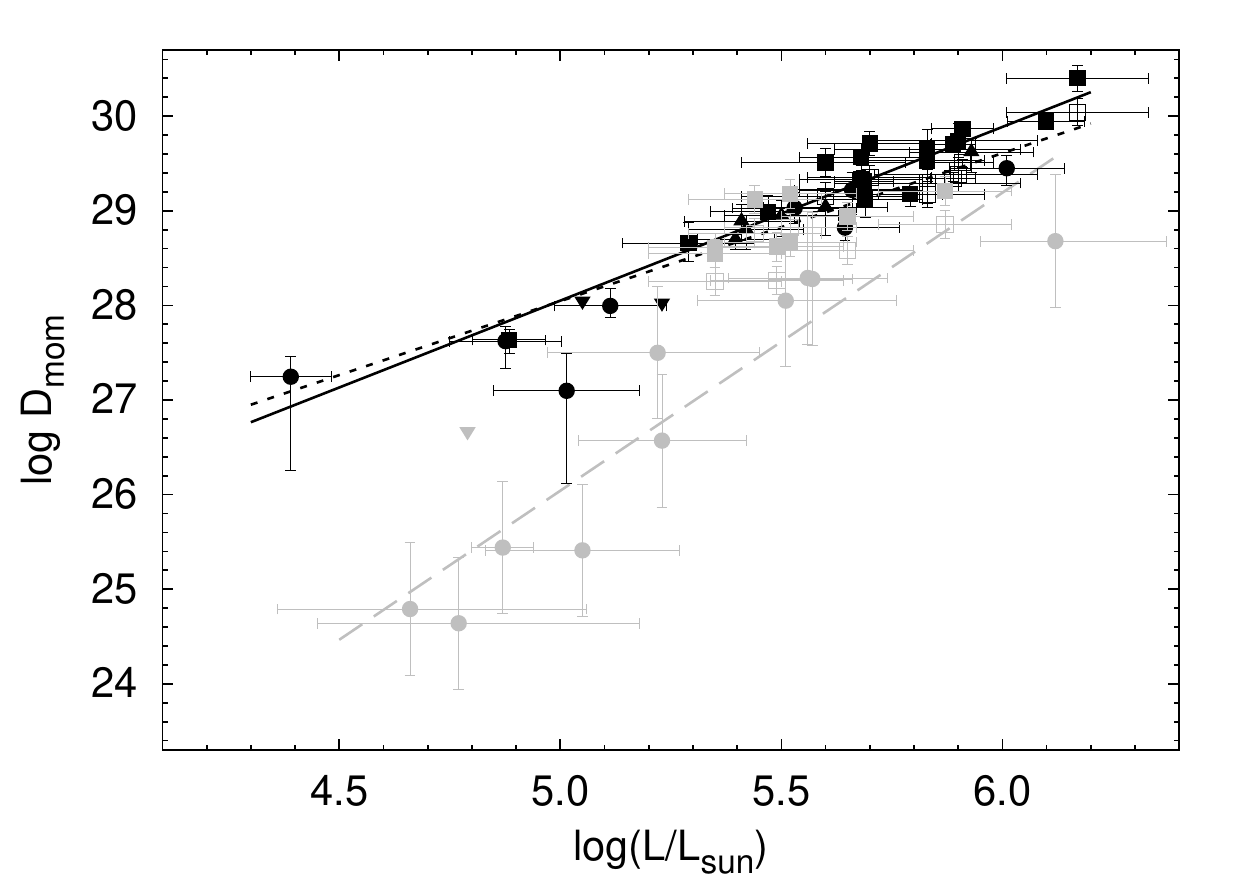}}
\caption{Modified wind-momentum -- luminosity distribution of
Galactic stars. Different luminosity classes are shown using
circles, triangles and squares for, respectively, V, III-II and I
class objects. Upper limits are shown as inverted
triangles. Black/grey symbols refer to \fastwind/\cmfgen\
analyses. Clumping corrected \Dmom\ values, using a clumping 
correction of $-0.37$~dex for objects with a \ha\ emission line
profile, are shown with open symbols. {\it Left:} stars analysed
by \cite{mokiem05} using an automated fitting method. The solid
line corresponds to the empirical wind-momentum luminosity relation
(WLR) for rates not corrected for clumping.  A dotted line is the
WLR for clumping corrected values. {\it Right:} total Galactic
sample. Black solid and dotted lines, respectively, correspond to
WLRs fitted to the complete sample using uncorrected and clumping
corrected rates. The grey dashed line is the WLR obtained from
fitting only the dwarfs analysed by \cite{martins05b}, adopting
their rates for $f = 1$ (grey circles). Upper limits were not
considered in determining any of the best fits.}
\label{fig:wlr-gal}
\end{center}
\end{figure*}

For the Galaxy we consider a sample based on the analyses performed by
\cite{repolust04}, \cite{mokiem05}, \cite{martins05b} and \cite{crowther06}.
Note that the second study includes a reanalysis of the Cyg~OB2 stars
analysed by \cite{herrero02}, and HD~15629 and $\zeta$~Oph studied by
\citeauthor{repolust04} The relevant atmospheric parameters adopting
non-clumped mass-loss rates are listed in Tab.~\ref{tab:wind_par_gal}. Two
entries are given for HD~15629 and HD~93250, as they were analysed
separately using \fastwind\ and \cmfgen. In the following determination of
the Galactic WLR, we adopt the results of the \fastwind\ studies for
reasons of consistency, but note that the differences of both analyses are
not signficant.

In Fig.~\ref{fig:wlr-gal} the distribution of the Galactic stars in
the modified wind-momentum vs.\ luminosity diagram is presented, where
black/grey symbols refer to \fastwind/\cmfgen\ analyses. Different
luminosity classes are distinguished using circles, triangles and
squares for, respectively, class V, III-II and I objects. 
The open symbols show \Dmom\ values resulting from scaling the \mdot\ by a
factor of 0.44, corresponding to $-$0.37 in $\log \Dmom$, for supergiants
exhibiting \ha\ emission line profiles. A reduction of this amount was
proposed by \cite{repolust04} to correct for the fact that these stars have
a systematically higher wind momentum compared to O dwarfs and
theoretical predictions \citep[see also][]{pulsIAUS212, markova04}. The
physical interpretation for this systematic offset proposed by these authors
is connected to wind clumping. Because \ha\ is a recombination line, its
strength scales with the (wind)density $\rho$ squared. In a uniformly
clumped wind, the emission will increase by a factor $f =
<\rho^2>/<\rho>^{2}$, where $f$ is referred to as the clumping factor. 
As the \ha\ line-forming region of stars with \ha\ in emission is more
extended compared to stars in which the profile is seen in absorption, 
these more extended regions must be more clumped than the innermost wind
regions of those stars with absorption type profiles. In this
interpretation, therefore, the observed offset suggests a (spatial) gradient
in the clumping factor or a difference in the clumping properties of thin
and thick winds. The (differential) clumping factor that corresponds to the
applied scaling is $f = 1/0.44^{2} \simeq 5$. We will return to the issue of
clumping in Sect.~\ref{sec:clumping}.

The left-hand side of Fig.~\ref{fig:wlr-gal} compares the stars that have
been analysed by \cite{mokiem05} in a homogeneous way by using an
automated fitting method. We have constructed a WLR by fitting a
power law, while accounting for both the symmetric errors in \lstar\
and the asymmetric errors in \Dmom, to the observed distribution. This
empirical WLR is shown as a solid and dotted black line for
uncorrected and clumping corrected wind momenta, respectively.  The
clumping corrected relation is found to be flatter because the
clumping corrections only affected the two brightest objects.
In Sect.~\ref{sec:mdot_vs_z} we will compare the empirical and
theoretical WLRs in the observed \lstar\ range.

On the right-hand side of Fig.~\ref{fig:wlr-gal} the observed \Dmom\
distribution for the complete Galactic sample is shown. The empirical
WLRs determined for this sample are again shown as a black solid and
dotted line for the uncorrected and clumping-corrected rates. As can
be seen in Tab.~\ref{tab:wlr-par} the fit coefficients for both
relations are in very good agreement with those derived from the
homogeneously analysed sample.

The Galactic sample comprises results from both \cmfgen\ and \fastwind\
studies. We note that \fastwind\ is so far limited to studies of the visual
spectral region -- therefore \ha\ is the most important \mdot\ diagnostics
-- while \cmfgen\ (also) uses the ultraviolet regime. In the latter approach
the UV lines are given more weight in the mass-loss determination. In the
case of weak winds ($\lesssim 10^{-7} \msunyr$) the \mdot\ determinations
rely almost exclusively on fits to \civ\,$\lambda\lambda 1548,1551$
\citep{martins05b}. 

Figure~\ref{fig:wlr-gal} shows good agreement between \cmfgen\ and
\fastwind\ studies for relatively high luminosities ($\log \lstar/\lsun
\gtrsim 5.5$), consequently high wind densities. For lower luminosities,
the UV analyses of the dwarf sample studied by \cite{martins05b} show a
systematic discrepancy with the average relations. This is emphasised by the
grey dashed line, which shows the average relation for these objects.
The fit coefficients of this latter relation are also listed in 
Tab.~\ref{tab:wlr-par} and clearly
signal a discrepency between UV based mass-loss determinations and
theoretical expectations. This is referred to as the ``weak wind problem''
\citep[see e.g.][]{bouret03}. We will discuss this further in
Sect.~\ref{sec:weakwinds}.

\begin{figure*} 
    \resizebox{18cm}{!}{
    \includegraphics{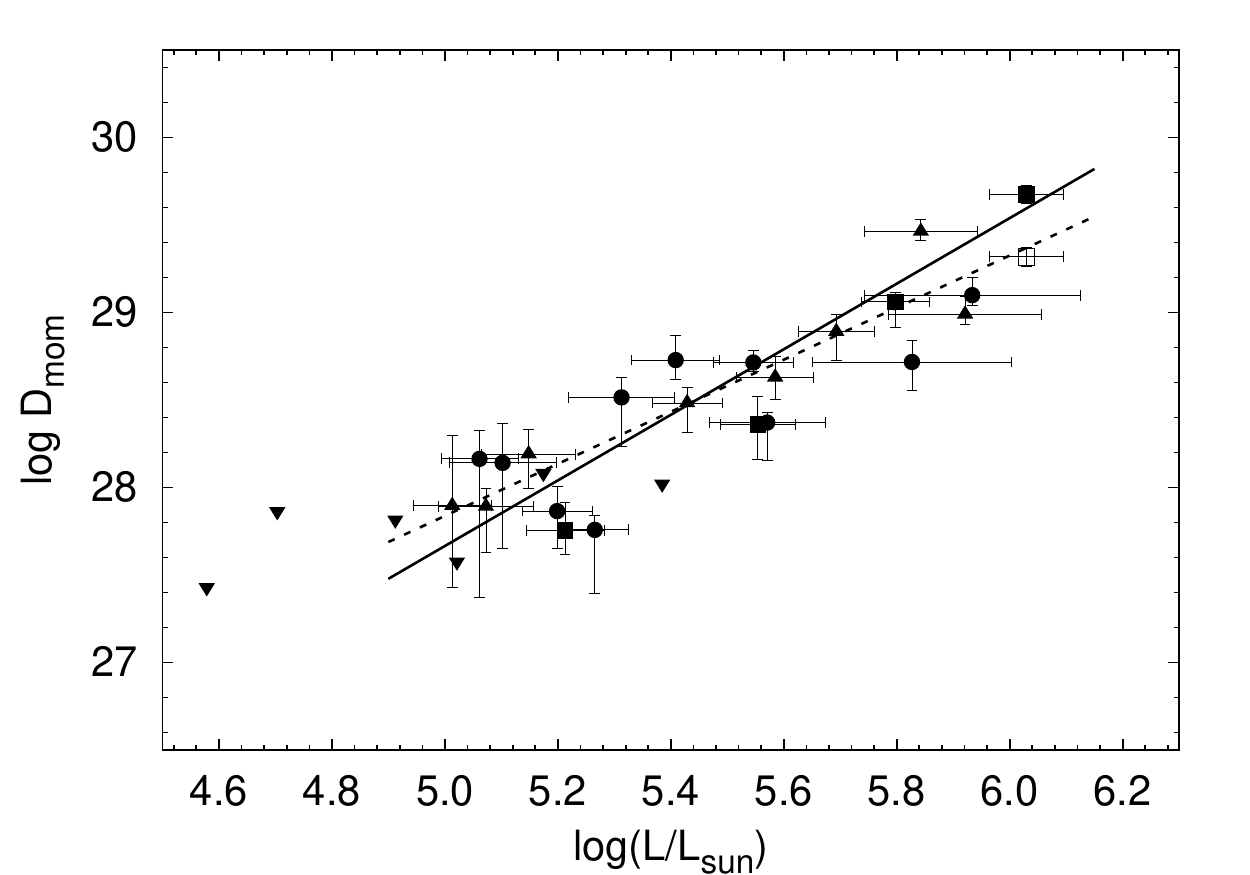} 
    \includegraphics{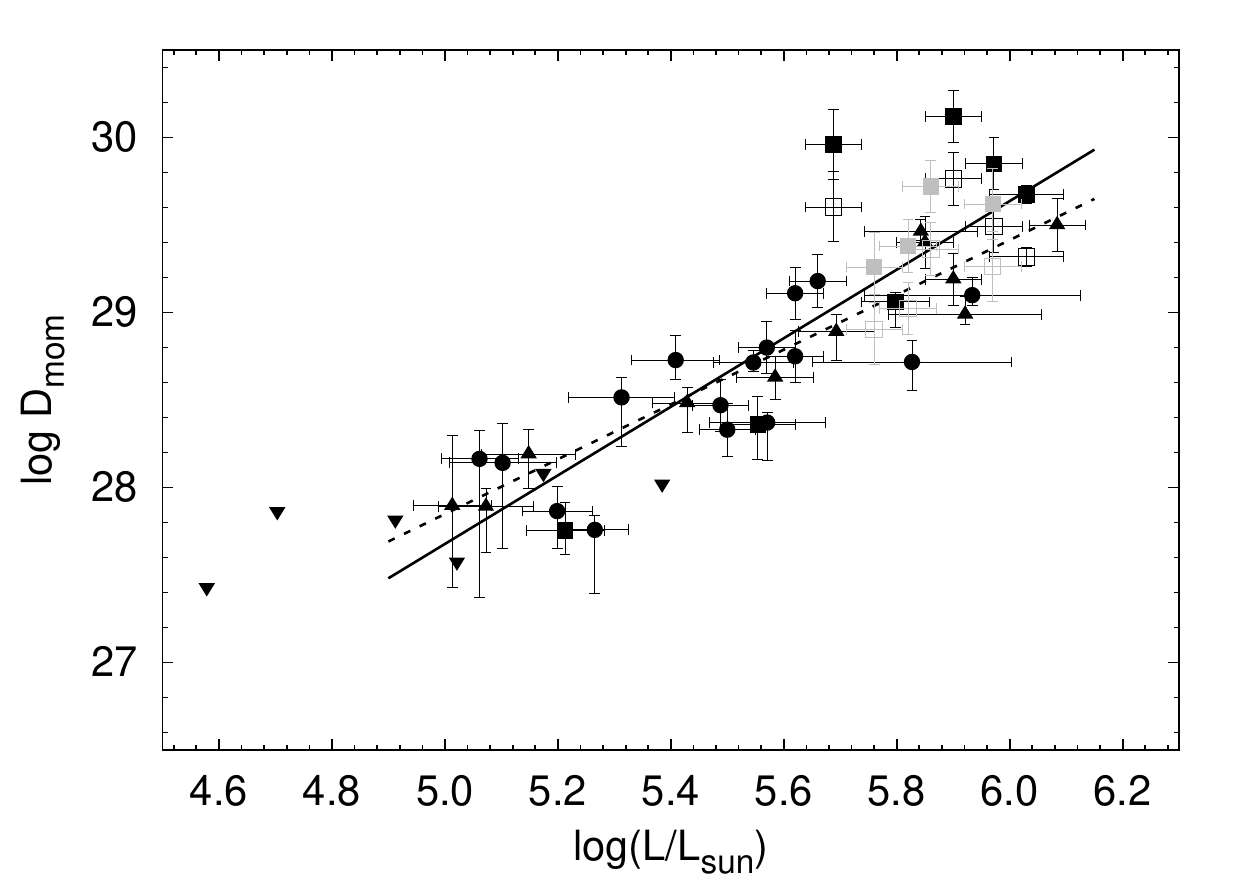} 
  }
  \caption{Modified wind momentum -- luminosity distribution for LMC
  stars. Symbols have the same meaning as in
  Fig.~\ref{fig:wlr-gal}. {\it Left:} stars analysed by
  \cite{mokiem07} using an automated fitting method. Dotted and
  dashed lines are empirical WLRs fitted to uncorrected and clumping
  corrected \Dmom\ values, respectively. {\it Right:} total LMC
  sample. Fitted empirical WLRs are shown as a solid and dashed line
  for uncorrected and clumping corrected wind momentum rates. The fit
  coefficients for the different WLR relations are given in
  Tab.~\ref{tab:wlr-par}. Upper limits were not considered in
  determining any of the best fits.}
   \label{fig:wlr-lmc}
\end{figure*}

\subsection{LMC}

The studies from which we have drawn the LMC sample are those of
\cite{crowther02}, \cite{massey04, massey05}, \cite{evans04b} and
\cite{mokiem07}. In Tab.~\ref{tab:wind_par_lmc} the relevant
atmospheric parameters are listed. Note that some of the objects in
this table have been analysed in more than one study. In those cases
we adopted the results from studies using the automated fitting
procedure developed by \cite{mokiem05}. If these were not available we
preferred results from studies that used both the optical and UV
spectral range over results that consider only the optical regime. For a
number of stars analysed by \cite{mokiem07}, no UV measurement of the
wind velocity was available. Consequently, these authors estimated
\vinf\ by scaling the escape velocity at the stellar surface (\vesc)
with a constant factor of 2.6 \cite[cf.][]{lamers95}. To account for
the metallicity dependence and to facilitate a comparison with
theoretical predictions, we accordingly rescaled this \vinf\ using
Eq.~\ref{eq:vinf_z} with $Z = 0.5\,\zsun$ and $n = 0.13$
\citep[cf.][]{leitherer92}. In Tab.~\ref{tab:wind_par_lmc} the
rescaled values are given between brackets. As \vinf\ also influences
the density in the line forming region of wind sensitive lines
(because of the requirement of mass continuity), also a rescaling of
\mdot\ was required. For this we used the wind-strength
parameter
\begin{equation}
   Q = \frac{\mdot}{R^{3/2}_\star \vinf}~.
\end{equation}
that conserves the \ha\ equivalent width
(\citealt{schmutz89}, see also \citealt{puls96} and
\citealt{dekoter97}).
The combined effect of these re-scalings is a reduction of the
modified-wind momentum by 0.08~dex.

The distribution of the modified-wind momenta as a function of stellar
luminosity for the LMC stars is shown in Fig.~\ref{fig:wlr-lmc} using
the same symbols as in Fig.~\ref{fig:wlr-gal}. On the left-hand side of the
figure we (again) only consider objects that have been analysed by
\cite{mokiem07} using an automated fitting method.  The solid and
dotted lines give the mean relations for uncorrected and clumping
corrected rates. The correction applied was the same as for the
Galactic stars. We note however that the metallicity dependence
of this correction is as yet unknown.

For the total sample, shown on the right-hand side of the
Fig.~\ref{fig:wlr-lmc}, the scatter is slightly larger. In particular
the bright supergiants Sk~$-67$~22 at $\log \lstar/\lsun \approx 5.7$
seems to stand out. Probably its discrepant position can be explained
by the fact that \cite{massey05} could only determine a lower limit
for its effective temperature, hence it could be intrinsically
brighter. 
Despite the increased scatter, the obtained WLRs are in good
agreement with the relations determined from the homogeneously
analysed sample. This can be seen from the fit coefficients given in
Tab.~\ref{tab:wlr-par}. For a confrontation of the empirical behaviour
of the WLR with predictions, we refer the reader to Sect.~\ref{sec:mdot_vs_z}.

\subsection{SMC}

\begin{figure*} 
  \resizebox{18cm}{!}{
    \includegraphics{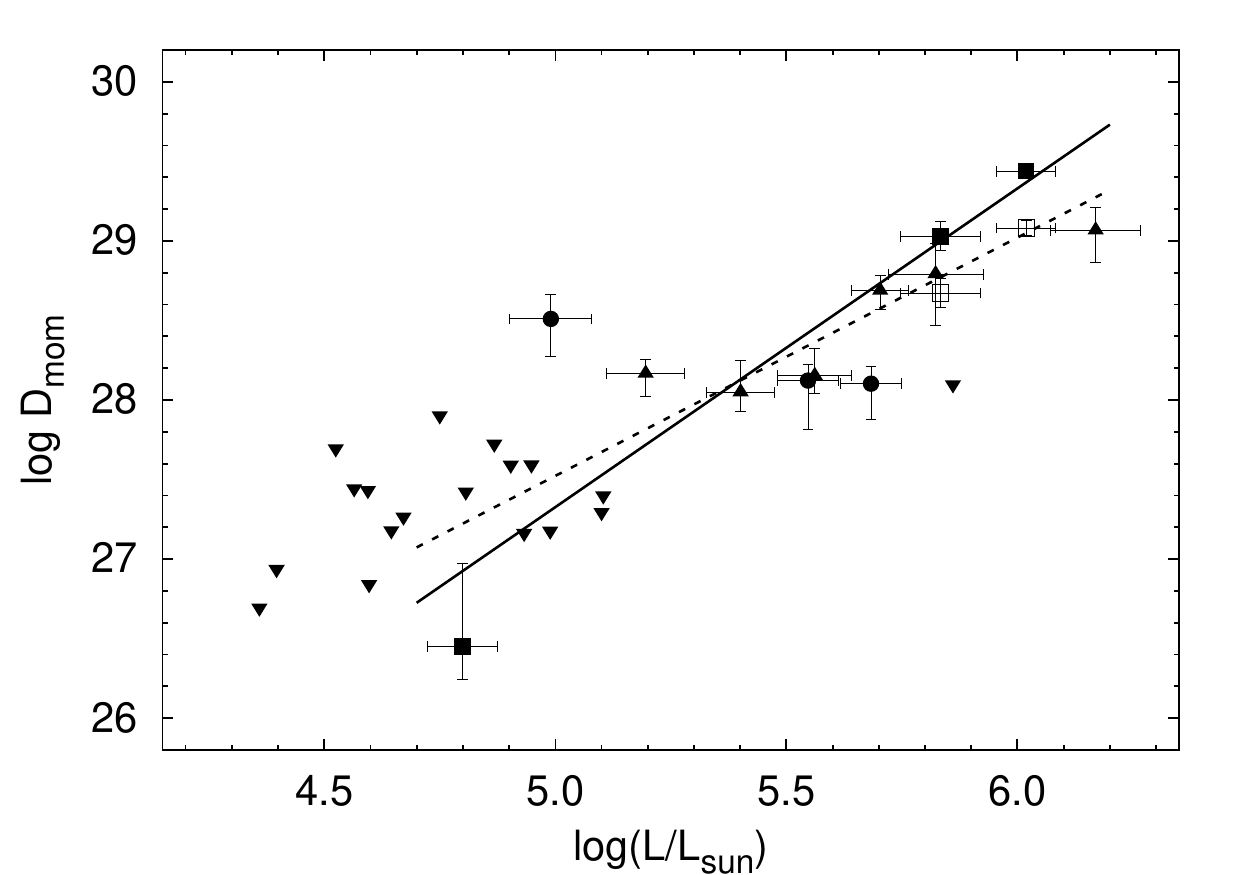} 
    \includegraphics{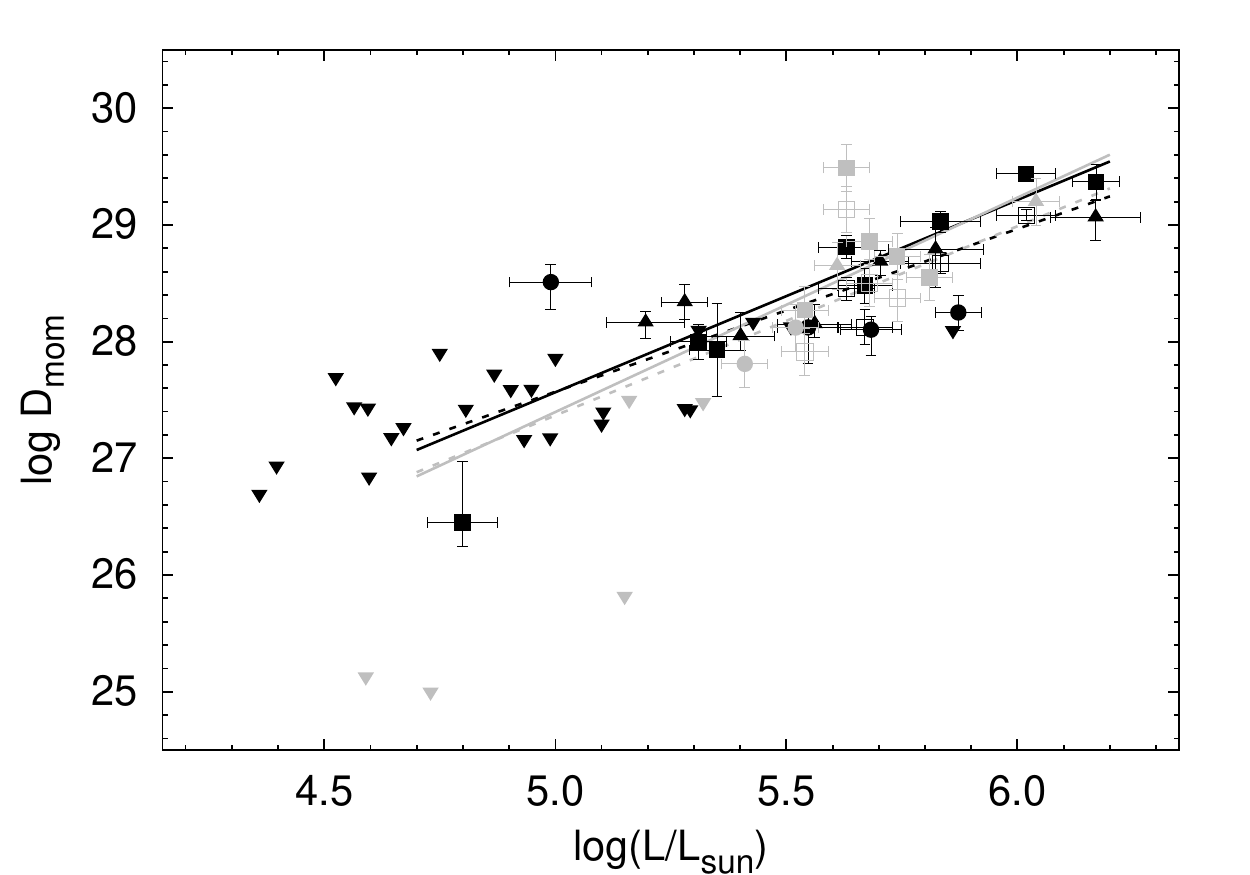} 
  }
  \caption{Modified wind momentum -- luminosity distribution for SMC
    stars. The symbols have the same meaning as in
    Fig.~\ref{fig:wlr-gal}. To construct the uncorrected and clumping
    corrected empirical WLR (solid and dashed line) the dwarf
    NGC346~033 at $\log \lstar/\lsun \approx 5.0$ was not taken into
    account as the scaling of the escape velocity resulted in an
    anomalously high wind velocity. In Tab.~\ref{tab:wlr-par} the fit
    coefficients for the individual relations are listed. {\it Left:}
    stars analysed by \cite{mokiem06} an automated fitting
    method. {\it Right:} total SMC sample. It turns out that the WLR
    is strongly influenced by the objects with $\log \lstar/\lsun <
    5.3$. This is shown by the grey lines, which correspond to the
    empirical WLRs calculated ignoring these objects. For establishing
    the empirical $\mdot(Z)$ as well as for the comparison with theory
    we will use the latter set of WLRs. Upper limits were not
    considered in determining any of the best fits.}
   \label{fig:wlr-smc}
\end{figure*}

In Tab.~\ref{tab:wind_par_smc} atmospheric parameters are given for
the SMC sample that was compiled from the studies of
\cite{hillier03}, \cite{bouret03},
\cite{trundle04}, \cite{massey04}, \cite{evans04b}, \cite{trundle05},
\cite{massey05} and \cite{mokiem06}. For objects that were fitted in
multiple studies we adopted results of one of these investigations
following the same rules as applied to the LMC stars.  Wind
velocities given between brackets correspond to values that were
calculated from the escape velocity at the stellar surface. These
velocities and the associated mass-loss rates were scaled in a similar
manner as was done for the LMC. Adopting $Z=0.2\,\zsun$, the \Dmom\
values were scaled down by 0.18~dex.

Figure~\ref{fig:wlr-smc} shows the distribution of the SMC stars in
the \Dmom\ vs.\ \lstar\ diagram. The symbols used are the same as in
Fig.~\ref{fig:wlr-gal}. On the left-hand side of the figure only the
objects that have been analysed using an automated fitting method by
\cite{mokiem06} are shown. For $\log \lstar/\lsun \lesssim 5.3$ the
majority of the mass-loss determinations are upper limits. The
empirical WLRs are shown using a solid and dotted line for uncorrected
and clumping corrected wind momenta. Note the position of the dwarf
NGC346-033 at $\log \lstar/\lsun \approx 5.0$. We did not include this
object in the fits, as its high wind momentum is the result of an
anomalously high wind velocity resulting from the scaling with \vesc\
\citep[also see][]{mokiem06}.

The right-hand side of Fig.~\ref{fig:wlr-smc} shows the wind momentum
distribution for the total SMC sample. For this large sample the low
luminosity part of the diagram still remains scarcely populated.  Also
note the upper limits determined using \cmfgen\ for $\log \lstar/\lsun
< 5.2$.
This could point to a steeper WLR relation for UV based mass-loss
estimates compared to those obtained from \ha, as was found for the
galactic case. However, this is not a firm statement as by far the
bulk of the SMC targets in this luminosity range only show upper
limits.

The empirical WLRs for the total sample are shown on the right-hand side
Fig.~\ref{fig:wlr-smc} as a black solid line for uncorrected wind momenta
and as a dotted line for clumping corrected values. We find that these
relations are strongly influenced by the objects with $\log
\lstar/\lsun < 5.3$. This is shown by the grey solid and dashed lines,
which correspond to the respective WLRs calculated ignoring these
objects. As the low luminosity part of the SMC diagram is rather
uncertain, we opted to use this latter set of relations. Note that
this choice will not influence our determination of the metallicity
dependence of \mdot, as this will be based on the stars brighter than
\lstar/\lsun = 5.3. In this range, the fit uncertainties are small (see
Sect.~\ref{sec:mdot_vs_z}).

\section{Mass loss versus metallicity}
\label{sec:mdot_vs_z}

Now that we have established the empirical modified-wind momentum
luminosity relation for the three galaxies, we can determine 
the $\mdot(Z)$ relation. Before doing so, we first
inter compare the results for these three environments, and confront
them with predictions.

\subsection{Global comparison}

\begin{figure*} 
  \resizebox{18cm}{!}{
    \includegraphics{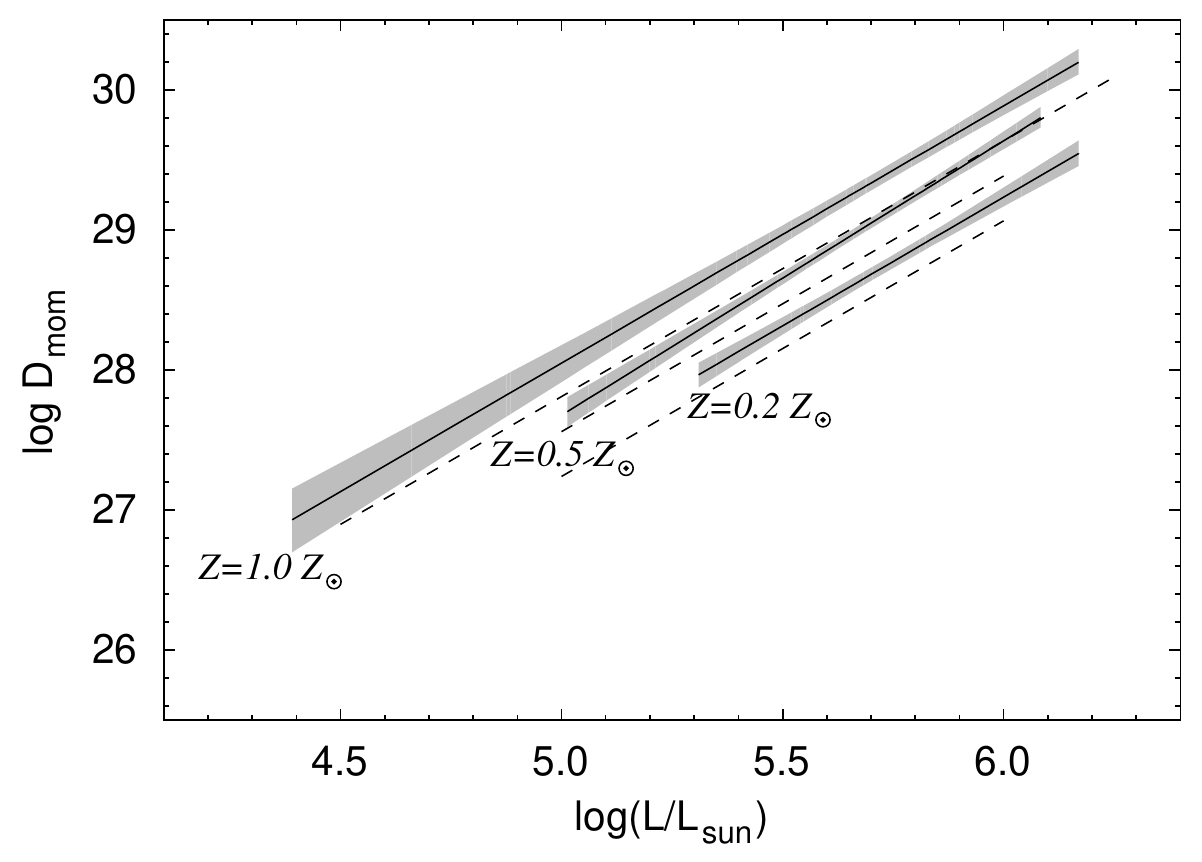} 
    \includegraphics{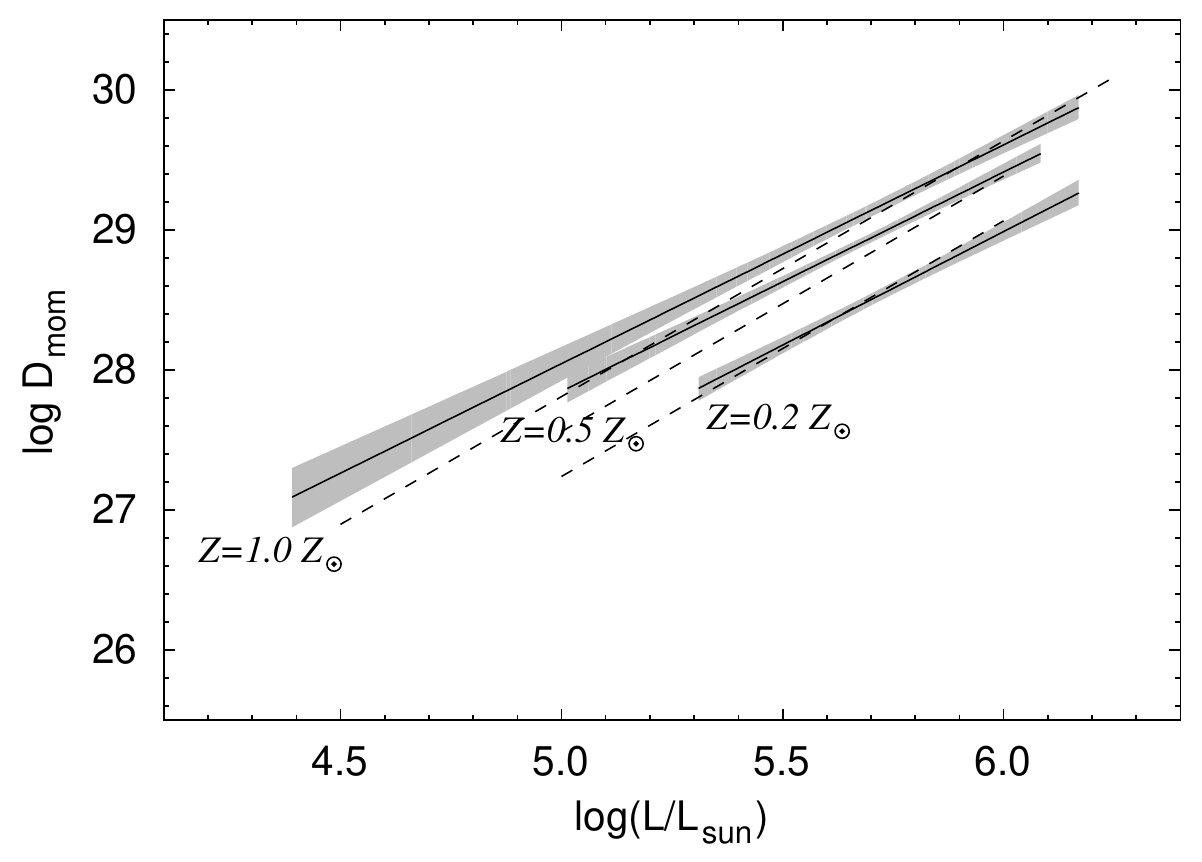} 
  }
  \caption{Comparison of the observed wind momentum -- luminosity
    relations (solid lines) with the predicted relations of
    \cite{vink00, vink01} (dotted lines). Top, middle and bottom lines
    of each line style, respectively, correspond to Galactic, LMC and
    SMC observed and predicted WLRs. Left and Right plots show,
    respectively, empirical WLRs including uncorrected and clumping
    corrected wind momentum rates. See Sect.~\ref{sec:gal} for the
    applied clumping correction. One sigma confidence intervals for
    the empirical relations are shown as grey areas.}

   \label{fig:wlr-glob}
\end{figure*}

In Fig.~\ref{fig:wlr-glob} the empirical modified-wind momentum
relations determined for the total observed samples (solid lines) are
shown alongside the predicted relations of \cite{vink00, vink01}
(dotted lines). The top, middle and bottom lines of each line style,
respectively, correspond to the Galactic, LMC and SMC observed and
predicted relations. To facilitate a meaningful comparison one sigma
confidence intervals for the observed WLRs are shown as grey
areas. First focusing on the left-hand side, showing the
relations without a clumping correction, we see that the 
empirical relations are clearly separated from each other beyond the
fitting uncertainties. {\em We interpret this as quantitative evidence
for a successive decrease of the mass-loss rates of massive stars in
the Galactic, LMC, and SMC environment.}

For the entire investigated luminosity range, the relative
separations of the empirical WLRs agree well with the
separations predicted by \cite{vink01}, although the empirical
WLRs do show a systematic offset compared to the theoretical
results. Focusing for a moment on the relative separations only:
measured at $\log \lstar/\lsun = 5.75$, which coincides with the
region where the fit uncertainties are the smallest, the offsets 
between the empirical and theoretical relations relative to the Galaxy
are: $0.28$ and $0.25$ for the LMC, and $0.62$ and $0.57$ for the
SMC.

The systematic offset between observations and theory is of the
order of 0.2~dex. This suggests that {\em if wind clumping does
affect the determination of empirical mass-loss rates, but would only
have a marginal effect on mass-loss predictions}, the empirical
mass-loss rates would be overestimated by at most a factor of $\sim$
two.

Indeed, when turning to the clumping corrected empirical relations, 
the relative offset almost disappears. This is shown on
the right-hand side of Fig.~\ref{fig:wlr-glob}, where we show the wind
momentum relations obtained from the total observed samples accounting
for this correction. Note that, strictly speaking, this agreement
pertains only to stars {\em not} showing \ha\ in emission, as the
clumping corrections is designed such as to scale the ones that do
show \ha\ emission to those that do not (see Sect.~\ref{sec:gal}).
 The slopes of the empirical relations, however,
are slightly flatter, as the clumping correction preferentially
affects the high-luminosity objects.

\subsection{The empirical $\mdot(Z)$ relation}
\label{sec:empirical-dmdtz}

To calculate the mass loss metallicity dependence we use the relative
separation between the empirical modified-wind momentum relations at
$\log \lstar/\lsun = 5.75$, for which the fit uncertainty in all three
relations is at its minimum. Note that the differential slopes also
allow for a luminosity dependent $Z$-dependence, however, based on the
uncertainties we doubt whether this would be meaningful. Moreover, for
the SMC it was shown that this slope is very sensitive to the rather
uncertain low luminosity domain. In Tab.~\ref{tab:wlr-par} the power
law indices for the observed $\mdot(Z)$ dependence for the LMC and SMC
relative to the Galaxy are listed. These were calculated using
Eq.~\ref{eq:z_dep} assuming $n=0.13$. The indices determined from the
homogeneously analysed sample are in good agreement with those
determined for the total sample. For the LMC relatively larger
differences are found compared to the SMC, which is a result of the
smaller metallicity difference of this system relative to our
galaxy. This also explains the larger error bars.

To determine the global metallicity dependence, we made a linear fit
regarding the logarithm of the differential (i.e. with respect to the Galaxy)
modified wind momentum and the log of the differential metallicity.
Assuming smooth
winds, we find:
\begin{equation}
  \mdot \propto Z^{0.83 \pm 0.16}~.   
  \label{eq:empirical-m}
\end{equation}

We emphasize that the error bar in $m$ is a fitting error of the mass-loss rate
and the metallicity only.
Several uncertainties add to the quoted error. These
include {\em i)} those related to the adopted luminosity at which we
gauge the relation; {\em ii)} errors related to the neglect of the
correlation between luminosity and modified wind momentum
\citep[see]{markova04}; {\em iii)} errors related to uncertainties in the
\vinf(\mdot) dependence, i.e. in the adopted value $n = 0.13$,
and {\em iv)} those related to possible systematic uncertainties in the
metallicities of the Magellanic Clouds. The motivation for gauging
at $\log \lstar/\lsun = 5.75$ is given above. Least squares fit analysis including
the correlation between $L$ and $D$ yield similar results for the
stars analysed using the automated fitting method, though for the
total sample the Galactic 
$\log D(\log L)$ relation is found to be somewhat steeper. Deviations in the Galactic 
slope are to be expected \citep{markova04}. However, at $\log \lstar/\lsun = 5.75$
the wind momenta are not found to be very different and result in
uncertainties $\Delta m \lesssim 0.10$ dex. Uncertainties in the
velocity vs. metallicity dependence are not very well known, however, 
they are likely to be small as well. By far the largest source of error in $m$ 
is that due to uncertainties in the metallicities of the Magellanic Clouds,
notably that of the SMC. To illustrate this, if the SMC metal content
is not $\Delta [Z/{\rm H}] \sim -0.7 \pm 0.1 $ but rather $-0.5 \pm 0.1$, the 
empirical value of $m$ will increase to $1.13 \pm 0.25$. 

Keeping this in mind, the derived exponent $m$ of the empirical mass-loss
luminosity relation as given in Eq.~\ref{eq:empirical-m} 
appears to be consistent with the predicted value
of $m_{\rm pred} = 0.69 \pm 0.10$ \citep{vink01} and 
$m_{\rm pred} = 0.67$ from the alternative investigation by \cite{krticka06}.

If we apply the earlier-mentioned clumping correction scaling of
$-$0.37 dex in $\log \Dmom$ for stars showing \ha\ emission, the
metallicity dependence is found to be:
\begin{equation}
  \mdot \propto Z^{0.72 \pm 0.15}~.    
\label{eq_clump}
\end{equation}

This result too is consistent with both theoretical predictions quoted
above.

Three remarks need to be made. First, we can only state that this
$\mdot(Z)$ relation holds for stars more luminous than
$\sim$$10^{5.2}$~\lsun. The reasons are the lack of firm
\mdot\ determinations for stars less luminous than this number and the
inconsistency between the \ha\ and UV wind lines mass-loss diagnostics
(see below). Note in particular that the slope of the WLR might depend
on the wind-density and/or metallicity itself (as it is predicted from
a close inspection of the line-strength distribution functions,
cf. \citealt{puls00}), though not to such an extent as indicated by
the weak wind problem (cf. Fig.~\ref{fig:wlr-gal}). In so far, it is
possible that the metallicity dependence for low-luminosity objects
deviates from our previous results.

Second, the absolute agreement in wind momentum only holds if the
winds of the O dwarfs are not significantly clumped. If they are, the
predictions (which do not yet account for clumping) may overestimate
the mass-loss rate and therefore \Dmom. Third, {\em if} O dwarf winds
are clumped, but the clumping does not depend on metal content or
mass-loss (i.e. $f \neq f(Z,\mdot(Z))$) then the derived $m$ will not
be affected, and, therefore, the agreement in the observed and
predicted mass loss -- metallicity scaling will remain preserved. To
rephrase this last point: irrespective of what the actual
\mdot\ values are, {\em if} clumping were independent of
metallicity and wind density, the derived {\em power-law index} is
anticipated to be correct -- even if the winds are significantly
clumped.

Interestingly, these analyses suggest that the systematic errors in
the slope of the mass-loss-metallicity relation, $m$, as introduced by
potential wind clumping may be less relevant than remaining
uncertainties in the metallicity determinations of the Magellanic
Clouds.

\section{Discussion}
\label{sec:disc}

\subsection{The weak wind problem}
\label{sec:weakwinds}

As mentioned before, for the majority of stars at lower luminosities
the ultraviolet resonance lines based mass-loss rates disagree strongly
with theoretical predictions. Still, there are few (though different) 
Galactic objects at similar, low luminosity which display a ``normal'' \ha\
mass-loss rate, which might point either to different physics or 
questions the validity of at least one of the applied diagnostical tools. 

Going from high to low luminosity, figures~\ref{fig:wlr-gal} and
\ref{fig:wlr-smc} show that this discrepancy appears at $\lstar
\approx 10^{5.25}\,\lsun$. The dwarf star sample recently analysed by
\cite{martins05b} using \cmfgen\ suggests progressively weaker
winds than are predicted by current theory. At $L \sim 10^{4.75}
\lsun$ this discrepancy has reached a factor 100.
Note that given the still limited number of stars
investigated at relatively low luminosity alternative descriptions
(compared to a steeper slope) of the WLR might also be
appropriate. For instance, the WLR may retain the same slope but jump
to a two order of magnitudes lower value of \Dmom\ at $L \lesssim
10^{5.25} \,\lsun$ compared to the predictions. In the SMC, the
discrepancy could be even worse, as UV analyses only yield upper
limits. Note that also the \ha\ analyses of SMC objects at $L \lesssim
10^{5.25} \,\lsun$ yield almost exclusively results as upper limits.
So, in principle, these need not be contradictory to the UV based
\mdot.

What could be the reason for this ``weak wind problem''?  There are
three directions in which one might look for causes: first, errors in
the \ha\ and/or UV analysis; second, missing physics or invalid
assumptions in the wind predictions, or, third, the nature of the
stars showing the weak wind problem is different from that of normal
OB stars (that do appear to follow theory). 
Here, we focus on the first possibility. For more thorough
discussions on this topic we refer to \cite{martins04, martins05b}
and \cite{dekoter06}.

Two important advantages of the \ha\ method are that the ionisation and
abundance of hydrogen are well known.
One cannot make this statement with similar confidence for the
elements responsible for the UV resonance lines. For the weak wind stars,
the prime \mdot\ diagnostic -- \civ\ as mentioned -- is only a trace ion,
i.e. it does not represent the dominant ionisation stage. 
This makes it is susceptible to
X-ray emission, thought to originate from shocks that may develop in the
wind outflow. \cite{martins05b} present test calculations estimating that
X-rays may decrease the \civ\ ionisation by an order of magnitude,
requiring an increase of the mass loss by the same amount to preserve the
line fit. With typical uncertainties in the carbon abundance of a factor of
two, this shows that the UV method is much more prone to errors than the
\ha\ method. This does not imply that \ha\ does not suffer from
uncertainties. Besides the ambiguity in \mdot\ introduced by the
unknown velocity law in weaker winds which has been discussed already,
continuum rectification issues and potential nebular emission limit the
applicability of the \ha\ diagnostic to stars with $\mdot \gtrsim 10^{-7}
\msunyr$. To reach this limit requires high signal-to-noise data of stars
with no appreciable nebular emission, such as is the case for three of the
five Galactic stars at $L \lesssim 10^{5.15} \,\lsun$ -- i.e.\ in the (UV
identified) weak wind regime. These galactic stars are $\zeta$\,Oph,
Cyg\,OB2\,\#2, and HD\,217086. For two other stars at these low luminosities,
$\tau$\,Sco and 10\,Lac, we also established the mass-loss rate, 
however, the error bars on these determinations are large. {\em For the 
three stars mentioned, the \ha\ derived rates do appear to be more or 
less consistent with predictions.}  Note that some of 
the \civ\ analyses account for (canonical) X-ray emission suggesting 
that the problem may not be fully explicable by uncertainties in the 
UV method.

So far the origin of the weak wind problem has not yet
been identified. We stress that because of the poor overlap between the
\ha\ and UV diagnostics, it is at present difficult to exclude the
possibility that the (major) cause is simply connected with
uncertainties in the wind structure predicted by the model atmospheres
that are needed to derive the empirical \mdot -- notably the ionisation
of species responsible for the UV resonance lines. Further spectroscopy
in the IR, particularly using Br$\alpha$ as an ideal mass-loss indicator also for
weak winds, might unravel at least part of the problem.

\subsection{Small-scale clumping}
\label{sec:clumping}

So far, we have not discussed the physical reality of the ``clumping''
correction applied in Sect.~\ref{sec:obs_wlr} to bring the WLR
relation of the supergiants with \ha\ emission line profiles in accord
with those of dwarfs. Is wind clumping observed in O stars?  And, is
clumping more important in the winds of supergiants? Clumping was 
studied in detail by \cite{hillier91} to explain the 
scattering wings of \heii\ lines of Wolf-Rayet stars. Although
these stars are known to have very dense winds, their \heii\
scattering wings are weak features.  Consequently, for the weaker O
star winds these diagnostics cannot be applied. Nonetheless,
direct observational evidence that the winds of O stars are also
clumped was provided by \cite{eversberg98} from time series
spectroscopy of \heii\,$\lambda 4686$ in $\zeta$\,Pup, and by
\cite{crowther02, hillier03, bouret03, massa03, martins04, martins05b,
bouret05}, and \cite{fullerton06} from analysis of UV (resonance) lines, 
partly in combination with optical lines.

The fitting of the ultraviolet lines, notably P\,{\footnotesize v}\,$\lambda
\lambda 1118,1128$, \ov\,$\lambda 1371$, and \niv\,$\lambda 1718$, is found
to be improved if a distance dependent clumping is introduced -- with clumps
starting to form above the sonic point and reaching maximum clumping in the
outer wind\footnote{In this respect ``outer'' implies the formation
region of the UV resonance lines, which is inside of the region
that contributes the bulk of the radio flux.}
-- and a void
interclump medium is assumed. In the above listed studies a total of 25
stars have been analysed using the \cmfgen\ code. Clumping corrections are
applied in roughly half the sample, including both dwarfs and supergiants --
for objects brighter than $10^{5.15} \lsun$.  The maximum clumping factors
that are derived from these studies are in the range 10 -- 100. Though
clumping corrections have turned out to be necessary for some stars but not
for others, this does not exclude the possibility that all stars are clumped
to a certain degree: Due to the fact that only few lines react {\it
differentially} on clumping and thus allow for distinct statements, it is
rather difficult to exclude the presence of clumping in a certain object at
the present state of knowledge unless the wavelength coverage is fairly
complete (extending from the UV to the IR), and unless a large number of
systematic investigations have been performed.

In addition to these purely spectroscopic investigations, \cite{puls06}
analyzed a sample of Galactic O-type (super)giants more recently, combining
\ha, infrared, mm and radio fluxes to derive constraints on {\it the radial
stratification} of the clumping factor. Since all diagnostics employed have
a $\rho^2$ dependence, only {\it relative} clumping factors could be
derived, normalized to the values in the outermost, radio-emitting region,
whereas {\it absolute} values for the clumping factors and thus the actual
mass-loss rates remained unconstrained.

In contrast to present hydrodynamical simulations of the line-driving
instability but in (partial) agreement with previous studies it was found that
for {\it denser} winds clumping starts fairly close to the sonic point,
reaches a certain maximum (mostly within $r < 2 \rstar$) and decreases
outwards, where the maximum, normalized clumping factors are of the order of
3 to 6. For {\it weaker} winds, on the other hand, the clumping factors in
the inner and outermost regions turned out to be similar.

This appears to be consistent with
findings from the linear polarimetry of Luminous Blue Variables \citep
[]{davies05}, where it was found that (i) clumping must start close to the
photosphere to reproduce the observed levels of polarimetry, and (ii) denser
winds yield a larger amount of linear polarisation \citep[see also][for WR
stars]{robert89}.

The results by \cite{puls06} could mean one of two things: either the
outer wind is significantly clumped whilst the inner wind is even more
strongly clumped (which would imply that the UV results calling for
large clumping factors are correct), or the outer wind is not
significantly clumped and the current mass-loss predictions are 
approximately correct.
The latter possibility, of course, would require
an explanation of the different UV results.

\subsection{Concluding remarks}

The first result we mention is that the empirical $\mdot(Z)$ relation
appears to be in 
agreement with theoretical predictions for
luminosities larger than $\sim$$10^{5.2}$~\lsun. Second, for lower
luminosities, UV based \mdot\ determinations for dwarfs seem to
indicate a breakdown of the theory, leading to discrepancies of up to
a factor $10^{2}$ for the lowest luminosity stars with a measurable mass
loss. We found that for the \ha\ based mass-loss rates of 
three galactic stars, $\zeta$\,Oph,
Cyg\,OB2\,\#2, and HD\,217086, do follow the predictions down to 
$\log \lstar/\lsun \approx 4.9$.
Their \mdot\ values are $\sim 2 \times 10^{-7}~\msunyr$
or less, which is so low that they may suffer from
uncertainties that are not reflected in the derived error bars,
for instance those associated with the uncertain velocity law
in low density winds and/or the unnoticed presence of minor amounts 
of nebular emission. Still, given these
findings, it appears premature to exclude the possibility that
the (major) cause of the breakdown between empirical and predicted
mass loss at low luminosity is connected to modeling uncertainties
associated with the empirical \mdot\ values.

To further probe the cause of the possible theoretical breakdown
at low luminosity, it is therefore necessary to study those stars for
which {\em both} the \ha\ and UV line indicators of mass loss can be
applied -- although unfortunately there is only a very limited
mass-loss regime for which this is possible. To this end, the use of
Br$\alpha$ as a mass-loss diagnostic \citep{lenorzer04, repolust05}
will prove to be fruitful as it will widen the overlap
between hydrogen and UV line analyses.  Regarding clumping, the
\pv\ resonance line seems a powerful diagnostic \citep{massa03,
fullerton06}. Also, it may be fruitful to focus on the question 
whether all O-type stars suffer from clumping. Concerning the effect
of clumping on the mass loss vs.\ metallicity relation, the following
important remark can be made: regardless what the actual \mdot\ values
are, if clumping is universal in O-type stars, and if it does not
depend on wind properties such as density or metallicity, then the
derived scaling (i.e.\ the power law index $m$) remains correct.

Finally, let us briefly contemplate some implications of O-type star
mass-loss rates that are lower by factors of 3 to 100 than so far
assumed. These implications branch out into at least two directions:
wind hydrodynamics and stellar evolution. For our understanding of the
intrinsic instabilities associated with the wind driving mechanism
\citep{owocki94} the requirement of clumping factors $f \sim 10-100$
in the \ha\ line forming regions of dwarf O-type stars, 
which do not extend far beyond the sonic point, seems to imply
that these density inhomogeneities have developed already in the
sub-sonic part of the flow.
As an example, the O5\,V star N11-051 in
the LMC has an \ha\ line forming region that extends from the
photosphere to a velocity of
13 \kmsec, which is about 0.6 times the sonic velocity. For other O-type
dwarfs, similar results are found. This is not anticipated by theory,
which predicts that the {\em onset} of line-driven instabilities is
only at about the sonic velocity (where, admittedly, the growth
rate of the instability is large). Implications for the wind driving
mechanism may also be severe, although it should be noted that
clumping has not yet been incorporated in mass-loss predictions.

Clumping may also impact on our understanding of massive
star evolution. Weaker stellar winds will cause less loss of angular
momentum. Consequently the stars will not spin down as rapidly as
currently thought. It may even be expected that most Galactic massive
stars retain their initial rotational velocity properties during the
entire main sequence \citep{meynet00, maeder01}.
For the brightest stars -- corresponding to initial masses of 60
\msun\ or more -- the integrated main-sequence mass loss could
drop dramatically. With the current rates they are expected to 
lose 20 to 40 percent of their mass in the H-burning phase. This 
could drop by a factor of three, requiring the objects to lose
ten to tens of solar masses by other means in order to evolve towards
the hot Wolf-Rayet phase \cite[]{smith06}.  This could
either imply an increase of the duration of the LBV phase by a factor of
two, assuming that the (unknown) mechanism thought to be responsible
for the eruptive mass loss in this phase is unaffected by clumping
issues \cite[see e.g.][]{humphreys94}, or alternatively that 
some LBVs explode before reaching the Wolf-Rayet phase 
\citep{kotak06,gal-yam07,smith07}.

\begin{acknowledgements}
  M.R.M.\ acknowledges financial support from the NWO Council for Physical
  Sciences. JSV acknowledges financial support from an RCUK Fellowship. JP,
  FN and AH acknowledge support from the Spanish MEC through project
  AYA2004-08271-CO2. S.J.S.\ acknowledges the European Heads of Research
  Councils and European Science Foundation EURYI (European Young
  Investigator) Awards scheme, supported by funds from the
  Participating Organisations of EURYI and the EC Sixth Framework
  Programme.  
\end{acknowledgements}

\bibliographystyle{aa}

\begin{small}
\bibliography{7545} 
\end{small}

\clearpage

\appendix
\onecolumn

\section{Parameters analysed samples}

\begin{table*}[h]
  \caption{Wind parameters of Galactic O- and early B-type stars.}
  \label{tab:wind_par_gal}
\begin{center}
  \begin{tabular}{lllccccccc}
  \hline\\[-9pt] \hline \\[-7pt]
  Star &  ST &  \teff & \rstar & $\log \lstar$ & \mdot & \vinf & $\log \Dmom$ & Ref.  \\[2pt]
     &    &  [kK] & [\rsun] & [\lsun] & [\msunyr] & [\kmsec] & [${\rm g\,cm\,s^{-2}}\,\rsun$]\\[1pt]
\hline\\[-9pt]
Cyg~OB2 \#7	 & O3~If$^*$		 &  45.8 & 14.2	 & 5.91	 & $9.98\cdot10^{-6}$	 & 3080	 & 29.867 & 2\\
Cyg~OB2 \#11	 & O5~If$^+$		 &  36.5 & 21.9	 & 5.89	 & $7.36\cdot10^{-6}$	 & 2300	 & 29.700 & 2\\
Cyg~OB2~\#8C	 & O5~If		 &  41.8 & 13.1	 & 5.69	 & $3.37\cdot10^{-6}$	 & 2650	 & 29.313 & 2\\
Cyg~OB2~\#8A	 & O5.5~I(f)		 &  38.2 & 25.5	 & 6.10	 & $1.04\cdot10^{-5}$	 & 2650	 & 29.943 & 2\\
Cyg~OB2~\#4	 & O7~III((f))	 &  34.9 & 13.8	 & 5.40	 & $8.39\cdot10^{-7}$	 & 2550	 & 28.698 & 2\\
Cyg~OB2~\#10	 & O9.5~I		 &  29.7 & 30.1	 & 5.79	 & $2.63\cdot10^{-6}$	 & 1650	 & 29.175 & 2\\
Cyg~OB2~\#2	 & B1~I		 &  28.7 & 11.1	 & 4.88	 & $1.63\cdot10^{-7}$	 & 1250	 & 27.635 & 2\\
10~Lac 		 & O9~V		 &  36.0 & 8.4	 & 5.01	 & $6.06\cdot10^{-8}$	 & 1140	 & 27.098 & 2\\
$\zeta$~Oph	 & O9~V		 &  32.1 & 8.9	 & 4.88	 & $1.43\cdot10^{-7}$	 & 1550	 & 27.621 & 2\\
$\tau$~Sco	 & B0.2~V		 &  31.9 & 5.0	 & 4.39	 & $6.14\cdot10^{-8}$	 & 2000	 & 27.245 & 2\\
HD~115842	 & B0.5Ia		 &  25.5 & 34.2	 & 5.65	 & $2.00\cdot10^{-6}$	 & 1180	 & 28.940 & 4\\
HD~122879	 & B0Ia		 &  28.0 & 24.4	 & 5.52	 & $3.00\cdot10^{-6}$	 & 1620	 & 29.180 & 4\\
HD~14947	 & O5~If+		 &  37.5 & 16.8	 & 5.70	 & $8.52\cdot10^{-6}$	 & 2350	 & 29.710 & 1\\
HD~152234	 & B0.5Ia(Nwk)	 &  26.0 & 42.4	 & 5.87	 & $2.70\cdot10^{-6}$	 & 1450	 & 29.210 & 4\\
HD~15558	 & O5~III(f)		 &  41.0 & 18.2	 & 5.93	 & $5.58\cdot10^{-6}$	 & 2800	 & 29.620 & 1\\
HD~15629	 & O5~V((f))		 &  42.0 & 12.4	 & 5.64	 & $9.28\cdot10^{-7}$	 & 3200	 & 28.822 & 2\\
HD~15629	 & O5~V((f))		 &  41.0 & 12.0	 & 5.56	 & $3.16\cdot10^{-7}$	 & 2800	 & 28.290 & 3\\
HD~18409	 & O9.7~Ib		 &  30.0 & 16.3	 & 5.29	 & $1.02\cdot10^{-6}$	 & 1750	 & 28.660 & 1\\
HD~190864	 & O6.5~III(f)	 &  37.0 & 12.3	 & 5.41	 & $1.39\cdot10^{-6}$	 & 2500	 & 28.890 & 1\\
HD~192639	 & O7~Ib(f)		 &  35.0 & 18.7	 & 5.68	 & $6.32\cdot10^{-6}$	 & 2150	 & 29.570 & 1\\
HD~193514	 & O7~Ib(f)	 	 &  34.5 & 19.3	 & 5.68	 & $3.48\cdot10^{-6}$	 & 2200	 & 29.330 & 1\\
HD~193682	 & O5~III(f)		 &  40.0 & 13.1	 & 5.60	 & $1.73\cdot10^{-6}$	 & 2800	 & 29.040 & 1\\
HD~203064	 & O7.5~III:n((f))	 &  34.5 & 15.7	 & 5.50	 & $1.41\cdot10^{-6}$	 & 2550	 & 28.950 & 1\\
HD~207198	 & O9~Ib		 &  33.0 & 16.6	 & 5.47	 & $1.79\cdot10^{-6}$	 & 2150	 & 28.990 & 1\\
HD~209975	 & O9.5~Ib		 &  32.0 & 22.9	 & 5.69	 & $2.15\cdot10^{-6}$	 & 2050	 & 29.120 & 1\\
HD~210809	 & O9~Iab		 &  31.5 & 21.2	 & 5.60	 & $5.30\cdot10^{-6}$	 & 2100	 & 29.510 & 1\\
HD~210839	 & O6~I(n)fp		 &  36.0 & 21.1	 & 5.83	 & $6.85\cdot10^{-6}$	 & 2250	 & 29.650 & 1\\
HD~217086	 & O7~Vn		 &  38.1 & 8.4	 & 5.11	 & $2.13\cdot10^{-7}$	 & 2550	 & 27.995 & 2\\
HD~24912	 & O7.5~III(n)((f))	 &  35.0 & 14.0	 & 5.42	 & $1.08\cdot10^{-6}$	 & 2450	 & 28.800 & 1\\
HD~303308	 & O4~V((f+))		 &  41.0 & 11.5	 & 5.53	 & $1.63\cdot10^{-6}$	 & 3100	 & 29.030 & 1\\
HD~30614	 & O9.5~Ia		 &  29.0 & 32.5	 & 5.83	 & $6.04\cdot10^{-6}$	 & 1550	 & 29.530 & 1\\
HD~34078	 & O9.5~V		 &  33.0 & 7.5	 & 4.77	 & $3.00\cdot10^{-10}$	 & 800	 & 24.640 & 3\\
HD~37128	 & B0Ia		 &  27.0 & 24.0	 & 5.44	 & $2.25\cdot10^{-6}$	 & 1910	 & 29.120 & 4\\
HD~38666	 & O9.5~V		 &  33.0 & 6.6	 & 4.66	 & $3.00\cdot10^{-10}$	 & 1200	 & 24.790 & 3\\
HD~38771	 & B0.5Ia		 &  26.5 & 22.2	 & 5.35	 & $9.00\cdot10^{-7}$	 & 1525	 & 28.610 & 4\\
HD~42088	 & O6.5~Vz		 &  38.0 & 9.6	 & 5.23	 & $1.00\cdot10^{-8}$	 & 1900	 & 26.570 & 3\\
HD~46202	 & O9~V		 &  33.0 & 8.4	 & 4.87	 & $1.30\cdot10^{-9}$	 & 1200	 & 25.440 & 3\\
HD~46223	 & O4~V((f+))		 &  41.5 & 11.9	 & 5.57	 & $3.16\cdot10^{-7}$	 & 2800	 & 28.280 & 3\\
HD~66811	 & O4~I(f)		 &  39.0 & 19.4	 & 5.90	 & $8.80\cdot10^{-6}$	 & 2250	 & 29.740 & 1\\
HD~91943	 & B0.7Ia		 &  24.5 & 26.3	 & 5.35	 & $7.50\cdot10^{-7}$	 & 1470	 & 28.550 & 4\\
HD~91969	 & B0Ia			 &  27.5 & 25.3	 & 5.52	 & $1.00\cdot10^{-6}$	 & 1470	 & 28.670 & 4\\
HD~93028	 & O9~V			 &  34.0 & 9.7	 & 5.05	 & $1.00\cdot10^{-9}$	 & 1300	 & 25.410 & 3\\
HD~93128	 & O3~V((f))			 &  46.5 & 10.4	 & 5.66	 & $2.64\cdot10^{-6}$	 & 3100	 & 29.220 & 1\\
HD~93129A	 & O2~If*			 &  42.5 & 22.5	 & 6.17	 & $2.63\cdot10^{-5}$	 & 3200	 & 30.400 & 1\\
HD~93146	 & O6.5~V((f))	 	 & 37.0	 & 10.0	 & 5.22	 & $5.62\cdot10^{-8}$	 & 2800	 & 27.500 & 3\\
HD~93204	 & O5~V((f))	 		 & 40.0	 & 11.9	 & 5.51	 & $1.78\cdot10^{-7}$	 & 2900	 & 28.050 & 3\\
HD~93250	 & O3.5~V((f+))	 	 & 46.0	 & 15.9	 & 6.01	 & $3.45\cdot10^{-6}$	 & 3250	 & 29.450 & 1\\
HD~93250	 & O3.5~V((f+))	 	 & 44.0	 & 19.9	 & 6.12	 & $5.62\cdot10^{-7}$	 & 3000	 & 28.680 & 3\\
HD~94909	 & B0Ia		 	 & 27.0	 & 25.5	 & 5.49	 & $2.00\cdot10^{-6}$	 & 1050	 & 28.620 & 4\\[1pt]
  \hline														  
  \end{tabular}														  
\end{center}														  
References: (1)~\cite{repolust04}; (2)~\cite{mokiem05}; (3)~\cite{martins05b}; (4)~\cite{crowther06}			  
\end{table*}

\begin{table*}[!t]
  \caption{Wind parameters of O- and early B-type stars in the
  LMC. Wind velocities given between brackets are calculated from the
  escape velocity at the stellar surface. Identifications: ``BI'' from
  \cite{brunet75}, ``LH'' from \cite{lucke72}, except LH~51-496, which
  is identified by \cite{garmany94}, ``N11'' from \cite{evans06},
  ``R136'' from \cite{hunter97} and \cite{massey98} and ``Sk'' from
  \cite{sanduleak70}.}
  \label{tab:wind_par_lmc}
\begin{center}
  \begin{tabular}{lllccccccc}
  \hline\\[-9pt] \hline \\[-7pt]
  Star &  ST &  \teff & \rstar & $\log \lstar$ & \mdot & \vinf & $\log \Dmom$ & Ref.  \\[2pt]
     &    & [kK] & [\rsun] & [\lsun] & [\msunyr] & [\kmsec] & [${\rm g\,cm\,s^{-2}}\,\rsun$]\\[1pt]
\hline\\[-9pt]
BI 237		  & O2 V((f*))	 & 53.2	 & 9.7	 & 5.83	 & $7.81\cdot10^{-7}$	 & 3400		 & 28.717 & 1\\
BI 253		  & O2 V((f*))	 & 53.8	 & 10.7	 & 5.93	 & $1.92\cdot10^{-6}$	 & 3180		 & 29.100 & 1\\
HD~2670952	  & O6~Iaf+		 & 33.5	 & 25.0	 & 5.86	 & $1.10\cdot10^{-5}$	 & 1520		 & 29.720 & 4\\
HD~269050	  & B0~Ia		 & 24.5	 & 42.2	 & 5.76	 & $3.20\cdot10^{-6}$	 & 1400		 & 29.260 & 5\\
HD~269896	  & ON9.7~Ia+		 & 27.5	 & 42.3	 & 5.97	 & $7.50\cdot10^{-6}$	 & 1350		 & 29.620 & 5\\
Lh101:W3-24	  & O3~V((f))		 & 48.0	 & 8.1	 & 5.50	 & $5.00\cdot10^{-7}$	 & 2400		 & 28.330 & 3\\
LH58-496	  & O5~V(f)		 & 42.0	 & 10.5	 & 5.49	 & $6.00\cdot10^{-7}$	 & 2400		 & 28.470 & 2\\
LH64-16		  & ON2~III(f*)	 & 54.5	 & 9.4	 & 5.85	 & $4.00\cdot10^{-6}$	 & 3250		 & 29.400 & 2\\
LH81:W28-23	  & O3.5~V((f+))	 & 47.5	 & 10.0	 & 5.66	 & $2.50\cdot10^{-6}$	 & 3050		 & 29.180 & 2\\
LH81:W28-5	  & O4~V((f+))	 & 46.0	 & 9.6	 & 5.57	 & $1.20\cdot10^{-6}$	 & 2700		 & 28.800 & 3\\
LH90:ST2-22	  & O3.5~III(f+)	 & 44.0	 & 18.9	 & 6.08	 & $4.50\cdot10^{-6}$	 & 2560		 & 29.500 & 2\\
N11-004		  & OC9.7 Ib		 & 31.6	 & 26.5	 & 5.80	 & $1.62\cdot10^{-6}$	 & [2182]	 & 29.061 & 1\\
N11-008		  & B0.7 Ia		 & 26.0	 & 29.6	 & 5.55	 & $4.53\cdot10^{-7}$	 & [1480]	 & 28.362 & 1\\
N11-026		  & O2 III(f*)	 & 53.3	 & 10.7	 & 5.92	 & $1.66\cdot10^{-6}$	 & [2848]	 & 28.989 & 1\\
N11-029		  & O9.7 Ib		 & 29.4	 & 15.7	 & 5.21	 & $1.58\cdot10^{-7}$	 & [1440]	 & 27.754 & 1\\
N11-031		  & ON2 III(f*)	 & 45.0	 & 13.7	 & 5.84	 & $3.88\cdot10^{-6}$	 & 3200		 & 29.462 & 1\\
N11-032		  & O7 II(f)		 & 35.2	 & 14.0	 & 5.43	 & $7.37\cdot10^{-7}$	 & [1752]	 & 28.483 & 1\\
N11-033		  & B0 IIIn		 & 27.2	 & 15.6	 & 5.07	 & $2.23\cdot10^{-7}$	 & [1404]	 & 27.891 & 1\\
N11-038		  & O5 II(f+)		 & 41.0	 & 14.0	 & 5.69	 & $1.38\cdot10^{-6}$	 & [2377]	 & 28.890 & 1\\
N11-042		  & B0 III		 & 30.2	 & 11.8	 & 5.01	 & $1.73\cdot10^{-7}$	 & [2108]	 & 27.896 & 1\\
N11-045		  & O9 III		 & 32.3	 & 12.0	 & 5.15	 & $5.01\cdot10^{-7}$	 & [1415]	 & 28.191 & 1\\
N11-051		  & O5 Vn((f))	 & 42.4	 & 8.4	 & 5.31	 & $9.27\cdot10^{-7}$	 & [1927]	 & 28.515 & 1\\
N11-058		  & O5.5 V((f))	 & 41.3	 & 8.4	 & 5.27	 & $1.39\cdot10^{-7}$	 & [2259]	 & 27.758 & 1\\
N11-060		  & O3 V((f*))	 & 45.7	 & 9.7	 & 5.57	 & $4.77\cdot10^{-7}$	 & [2502]	 & 28.371 & 1\\
N11-061		  & O9 V		 & 33.6	 & 11.7	 & 5.20	 & $1.96\cdot10^{-7}$	 & [1734]	 & 27.865 & 1\\
N11-066		  & O7 V((f))		 & 39.3	 & 7.7	 & 5.10	 & $3.73\cdot10^{-7}$	 & [2116]	 & 28.140 & 1\\
N11-068		  & O7 V((f))		 & 39.9	 & 7.1	 & 5.06	 & $3.13\cdot10^{-7}$	 & [2769]	 & 28.164 & 1\\
R136-014	  & O3.5~If*		 & 38.0	 & 21.1	 & 5.90	 & $2.30\cdot10^{-5}$	 & 2000		 & 30.120 & 2\\
R136-018	  & O3~III		 & 45.0	 & 14.7	 & 5.90	 & $2.00\cdot10^{-6}$	 & 3200		 & 29.190 & 2\\
R136-033	  & O3~V		 & 47.0	 & 9.8	 & 5.62	 & $2.00\cdot10^{-6}$	 & 3250		 & 29.110 & 2\\
R136-055	  & O3~V		 & 47.5	 & 9.4	 & 5.62	 & $9.00\cdot10^{-7}$	 & 3250		 & 28.750 & 3\\
Sk~$-65$~47	  & O4~If		 & 40.0	 & 20.1	 & 5.97	 & $1.20\cdot10^{-5}$	 & 2100		 & 29.850 & 2\\
Sk~$-66$~169	  & O9.7~Ia+		 & 26.0	 & 40.0	 & 5.82	 & $6.00\cdot10^{-6}$	 & 1000		 & 29.380 & 4\\
Sk~$-67$~22	  & O2~If*		 & 42.0	 & 13.2	 & 5.69	 & $1.50\cdot10^{-5}$	 & 2650		 & 29.960 & 2\\
Sk~$-66$~100	  & O6 II(f)		 & 39.0	 & 13.6	 & 5.58	 & $8.81\cdot10^{-7}$	 & 2075		 & 28.629 & 1\\
Sk~$-66$~18	  & O6 V((f))		 & 40.2	 & 12.2	 & 5.55	 & $1.07\cdot10^{-6}$	 & 2200		 & 28.714 & 1\\
Sk~$-67$~166	  & O4 Iaf+		 & 40.3	 & 21.3	 & 6.03	 & $9.28\cdot10^{-6}$	 & 1750		 & 29.675 & 1\\
Sk~$-70$~69	  & O5 V		 & 43.2	 & 9.0	 & 5.41	 & $1.03\cdot10^{-6}$	 & 2750		 & 28.728 & 1\\[1pt]
  \hline
  \end{tabular}
\end{center}
References: (1)~\cite{mokiem07}, (2)~\cite{massey05}, (3)~\cite{massey04}, (4)~\cite{crowther02}, (5)~\cite{evans04b}
\end{table*}

\begin{table*}[!t]
  \caption{Wind parameters of O- and early B-type stars in the SMC.
  Wind velocities given between brackets are calculated from the
  escape velocity at the stellar surface. Identifications: ``AzV''
  from \cite{azzopardi75, azzopardi82}, ``MPG'' from \cite{massey89}
  and ``NGC330'' and ``NGC346'' from \cite{evans06}.}
\begin{center}
  \begin{tabular}{llccccccc}
  \hline\\[-9pt] \hline \\[-7pt]
  Star & ST &  \teff & \rstar & $\log \lstar$ & \mdot & \vinf & $\log \Dmom$ & Ref.  \\[2pt]
     &     & [kK] & [\rsun] & [\lsun] & [\msunyr] & [\kmsec] & [${\rm g\,cm\,s^{-2}}\,\rsun$]\\[1pt]
\hline\\[-9pt]
NGC346-001	  & O7~Iaf+	 & 34.1	 & 29.3	 & 6.02	 & $6.04\cdot10^{-6}$	 & 1330	 & 29.438 & 1\\
NGC346-007	  & O4~V((f))	 & 41.5	 & 11.1	 & 5.52	 & $2.70\cdot10^{-7}$	 & 2300	 & 28.120 & 3\\
NGC346-010	  & O7~IIIn((f))	 & 35.9	 & 10.2	 & 5.20	 & $4.88\cdot10^{-7}$	 & [1486]& 28.166 & 1\\
NGC346-012	  & B1~Ib	 & 26.3	 & 12.1	 & 4.80	 & $1.01\cdot10^{-8}$	 & [1272]& 26.450 & 1\\
NGC346-033	  & O8~V		 & 39.9	 & 6.6	 & 4.99	 & $6.02\cdot10^{-7}$	 & [3328]& 28.510 & 1\\
NGC330-013	  & O8.5~III((f)) & 34.5	 & 14.1	 & 5.40	 & $2.96\cdot10^{-7}$	 & 1600	 & 28.049 & 1\\
AzV~15		  & O7~II	 & 39.4	 & 18.3	 & 5.82	 & $1.12\cdot10^{-6}$	 & 2125	 & 28.791 & 1\\
AzV~26		  & O7~III	 & 40.1	 & 25.2	 & 6.17	 & $1.71\cdot10^{-6}$	 & 2150	 & 29.066 & 1\\
AzV~69		  & OC7.5~III((f))& 33.9	 & 18.6	 & 5.61	 & $9.20\cdot10^{-7}$	 & 1800	 & 28.650 & 5\\
AzV~70		  & O9.5~Ibw	 & 28.5	 & 28.4	 & 5.68	 & $1.50\cdot10^{-6}$	 & 1450	 & 28.860 & 2\\
AzV~75		  & O5.5~I(f)	 & 40.0	 & 25.4	 & 6.17	 & $3.50\cdot10^{-6}$	 & 2100	 & 29.370 & 4\\
AzV~83		  & O7~Iaf+	 & 32.8	 & 18.3	 & 5.54	 & $2.00\cdot10^{-6}$	 & 940	 & 28.270 & 5\\
AzV~95		  & O7~III	 & 38.2	 & 13.8	 & 5.56	 & $3.56\cdot10^{-7}$	 & 1700	 & 28.151 & 1\\
AzV~104		  & B0.5 Ia	 & 27.5	 & 20.0	 & 5.31	 & $3.24\cdot10^{-7}$	 & [1087]& 27.998 & 6\\
AzV~215		  & BN0 Ia	 & 27.0	 & 30.0	 & 5.63	 & $1.35\cdot10^{-6}$	 & 1400	 & 28.810 & 6\\
AzV~235		  & B0~Iaw	 & 24.5	 & 36.2	 & 5.63	 & $5.80\cdot10^{-6}$	 & 1400	 & 29.490 & 2\\
AzV~242		  & B1 Ia	 & 25.0	 & 36.6	 & 5.67	 & $8.40\cdot10^{-7}$	 & 950	 & 28.480 & 7\\
AzV~243		  & O6~V		 & 42.6	 & 12.8	 & 5.68	 & $2.64\cdot10^{-7}$	 & 2125	 & 28.102 & 1\\
AzV~296		  & O7.5~V((f))	 & 35.0	 & 11.9	 & 5.28	 & $5.00\cdot10^{-7}$	 & 2000	 & 28.340 & 4\\
AzV~372		  & O9~Iabw	 & 31.0	 & 28.7	 & 5.83	 & $2.04\cdot10^{-6}$	 & 1550	 & 29.028 & 1\\
AzV~388		  & O4~V		 & 43.3	 & 10.6	 & 5.55	 & $3.34\cdot10^{-7}$	 & 1935	 & 28.122 & 1\\
AzV~420		  & B0.5 Ia	 & 27.0	 & 21.7	 & 5.35	 & $2.76\cdot10^{-7}$	 & [1063]& 27.928 & 7\\
AzV~435		  & O3~V((f*))	 & 45.0	 & 14.2	 & 5.87	 & $5.00\cdot10^{-7}$	 & 1500	 & 28.250 & 8\\
AzV~456		  & O9.5~Ibw	 & 29.5	 & 30.6	 & 5.81	 & $7.00\cdot10^{-7}$	 & 1450	 & 28.550 & 2\\
AzV~469		  & O8.5~II((f))	 & 34.0	 & 20.6	 & 5.70	 & $1.10\cdot10^{-6}$	 & 1550	 & 28.688 & 1\\
AzV~488		  & B0.5~Iaw	 & 27.5	 & 32.6	 & 5.74	 & $1.20\cdot10^{-6}$	 & 1250	 & 28.730 & 2\\
MPG~355		  & O2~III(f*)	 & 52.5	 & 12.7	 & 6.04	 & $2.50\cdot10^{-6}$	 & 2800	 & 29.200 & 3\\
MPG~368		  & O4-5~V((f))	 & 40.0	 & 10.6	 & 5.41	 & $1.50\cdot10^{-7}$	 & 2100	 & 27.810 & 3\\[1pt]
  \hline
  \end{tabular}
\end{center}
References: (1)~\cite{mokiem06}, (2)~\cite{evans04b}, (3)~\cite{bouret03}, (4)~\cite{massey04},\\
            (5)~\cite{hillier03}, (6)~\cite{trundle04}, (7)~\cite{trundle05}, (8)~\cite{massey05}
\label{tab:wind_par_smc}
\end{table*}

\end{document}